\documentclass[preprint,12pt]{elsarticle}

\pdfoutput=1

\usepackage{graphicx}
\usepackage{cases}

\usepackage{newtxtext}

\usepackage{amsthm,bm}

\newcommand{\Pn}{P}
\newcommand{\Qm}{Q_\Gamma}

\newcommand{\ds}{\displaystyle}

\usepackage{stmaryrd} 
\newcommand{\jump}[1]{\left\llbracket #1 \right\rrbracket}

\newcommand{\nT}{n_\textup{\tiny W}}
\newcommand{\nJ}{{n_\textup{\tiny J}}}
\newcommand{\nC}{{n_\textup{\tiny V}}}

\newcommand{\bva}{{\bf V}}
 
\newcommand{\grad}{{\bf \nabla}}
\newcommand{\bn}{{\bf n}}

\newcommand{\bU}{{\bf u}}
\newcommand{\bW}{{\bf w}}
\newcommand{\bV}{{\bf v}}
\newcommand{\beq}{\begin{equation}}
\newcommand{\eeq}{\end{equation}}
\newcommand{\mT}{{\bf T}}
\newcommand{\mG}{{\bf G}}
\newcommand{\fu}{\textup{Func}}

\newcommand{\dr}[2]{\frac{\partial{ #1} }{\partial{ #2 }}}
\newcommand{\grandO}{{\cal O}(\Delta x^{k+1})}

\newcommand{\mE}{{\bf E}}
\newcommand{\mF}{{\bf F}}

\newcommand{\pr}{p^\textup{\scriptsize r}}

\usepackage{color}
\definecolor{rouge}{rgb}{1,0,0}
\definecolor{bleu}{rgb}{0,0,1}
\definecolor{vert}{rgb}{0,0.5,0}

\graphicspath{{Figures/}}

\begin{document}

\begin{frontmatter}

\title{Numerical modeling of the acoustic wave propagation across an homogenized rigid microstructure in the time domain}

\author[LMA]{Bruno Lombard}\corref{cor1}
\ead{lombard@lma.cnrs-mrs.fr}
\author[LANGEVIN]{Agn\`es Maurel}
\ead{agnes.maurel@espci.fr}
\author[LMS]{Jean-Jacques Marigo}
\ead{marigo@lms.polytechnique.fr}
\cortext[cor1]{Corresponding author. Tel.: +33 491 84 52 42 53.}
\address[LMA]{Laboratoire de M\'ecanique et d'Acoustique, CNRS UPR 7051, Aix-Marseille Universit\'e, Ecole Centrale Marseille, 13453 Marseille, France}
\address[LANGEVIN]{Institut Langevin, CNRS UMR 7587, ESPCI, 75005 Paris, France}
\address[LMS]{Laboratoire de M\'ecanique des Solides, CNRS UMR 7649, Ecole Polytechnique, 91120 Palaiseau, France}

\begin{abstract}
Homogenization of a thin micro-structure yields effective jump conditions that incorporate the geometrical features of the scatterers. These jump conditions apply across a thin but nonzero thickness interface whose interior is disregarded.   This paper aims  (i) to propose a numerical method able to handle the  jump conditions in order to simulate the homogenized problem in the time domain, (ii) to inspect the validity of the homogenized problem when compared to the real one. For this purpose,  we adapt an immersed interface method originally developed for standard jump conditions across  a zero-thickness interface. Doing so allows us to handle arbitrary-shaped interfaces on a Cartesian grid with  the same   efficiency and  accuracy of the numerical scheme than those obtained  in an homogeneous medium. Numerical experiments  are performed to test the properties of the numerical method and to inspect the validity of the homogenization problem. 
\end{abstract}


\end{frontmatter}



\section{Introduction}\label{SecIntro}

\begin{figure}[h]
\centering
\includegraphics[width=.4\columnwidth]{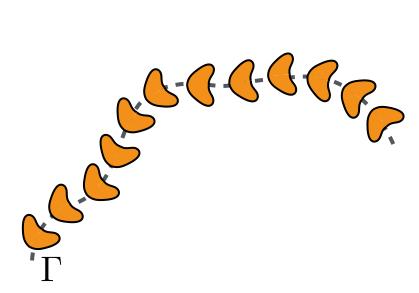} \hspace{.4cm}
\includegraphics[width=.4\columnwidth]{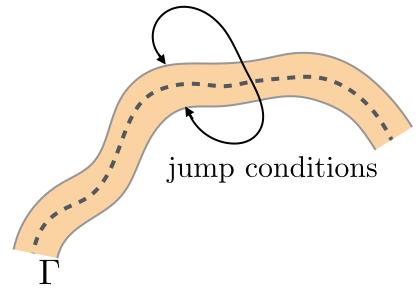} \\

\small (a) \hspace{4cm} (b)
\caption{ (a) The real problem involves a  structured film  composed of a periodic set of sound-hard scatterers in a homogeneous fluid, (b) the equivalent homogenized problem involves a thick interface across which jump conditions apply. The wavefield, on which the jump conditions apply,  is not represented. } 
\label{Fig1}
\end{figure}

The description of the interaction of waves with many scatterers of size much smaller than the wavelength is in principle simple since the scattering  is weak, and several approximated methods can be applied, owing to a small parameter being the scattering strength. In the case of many scatterers located periodically, homogenization techniques are well adapted to handle the problem within a rigorous mathematical framework. The classical homogenization of massive media predicts that scatterers occupying a large area can be replaced by an equivalent homogeneous and in general anisotropic medium occupying the same area \cite{Cioranescu}. The case of scatterers  located periodically along a mean line $\Gamma$ (Fig. \ref{Fig1}(a)) has been less regarded but it is nowadays accepted  that jump conditions across an equivalent interface have to be thought  (Fig. \ref{Fig1}(b)). Such jump conditions can be established using homogenization techniques which basically rely on the same ingredients than the classical homogenization of massive media, see {\em e.g.} \cite{Capdeville1,Capdeville2,Delourme1,Delourme2}; we call them interface homogenizations. 
 
 When the numerical resolution of the  problem is sought, such  equivalent media are of particular interest since they avoid to mesh the very small scales which are essential in the real problem. Indeed, these small scales  define  the scales of variation of the evanescent near wavefield excited in the vicinity of the scatterers. This is particularly true for scatterers located along a line since the  resulting film is dominated by the effect of the near field, or in other words by boundary layer effects. In the case of sound hard scatterers,  such interface homogenization has been proposed recently in \cite{Marigo16a,Marigo16c,Marigo16b}. These works follow those developed  in solid mechanics \cite{Marigo11} and they are adapted to the wave equation in the harmonic regime \cite{Marigo16a,Marigo16c} and in the time domain \cite{Marigo16b}. 

The goal of the present paper is twofold. First, we propose an accurate time-domain numerical scheme for the homogenized problem, thus   incorporating the  jump conditions across the equivalent thick interface. Next, we inspect the validity of the homogenized result;  notably we  exemplify the robustness of the jump conditions with respect to the ratio of the array spacing with the typical wavelength and with respect to a possible  curvature of the line $\Gamma$. 
Concerning the proposed numerical scheme, specific  aspects will be addressed:   
\begin{itemize}
\item the capability to  handle arbitrary-shaped interfaces on a Cartesian grid, without introducing spurious diffractions due to a naive stair-step discretization of the interfaces;
\item  the accuracy of the numerical scheme, despite the non-smoothness of the solution across the thick interface;
\item { the performances of the scheme in terms of computational cost; specifically, the scheme has to guaranty  an additional cost in the homogenized problem
(due to the treatment of the  jump conditions) which is highly  negligible compared with the computational cost in the real problem.}
\end{itemize}
To do so, a good strategy relies on an immersed interface method, originally developed in Refs. \cite{Li94,Zhang97}, and adapted to mechanical wave problems  \cite{Lombard04,Lombard06,Lombard08,Chiavassa11}. 
The extension of the method  to the present homogenized problem requires two ingredients. First, it requires to implement a generalized version  of the usual boundary conditions at an interface, able to  involve both the field and its spatial derivatives. Second, it must handle an interface with a non-zero thickness: the  values of the solution on both sides of the interface are linked together (via the jump conditions), whereas no field is computed inside the thin interface.

The paper is organized as follows. In section \ref{SecPhys}, the actual and the homogenized problems are  presented. 
Section \ref{SecNum} details the numerical methodology: a fourth-order ADER scheme combined with an immersed interface method, this latter constituting the core of the work. Section \ref{SecExp} presents numerical experiments. Comparisons with exact solutions confirm the efficiency and the  accuracy of the numerical modeling.  Next, comparisons with direct simulations for a real microstructure confirm the second-order accuracy of the effective model. Lastly, some perspectives are drawn in section \ref{SecCon}.


\section{The real problem and its homogenized version}\label{SecPhys}
The real problem concerns the propagation of acoustic waves through sound hard scatterers located periodically onto a mean line $\Gamma$ (Fig. \ref{Fig1}(a)). Periodically means a constant spacing $h$ between two scatterers  defined by the arc length along $\Gamma$. 
The resulting curved array is  surrounded by  a fluid with mass density $\rho$ and  isentropic compressibility $\chi$. In the fluid, the linearized 
Euler equations apply and Neumann boundary conditions for the  pressure
 apply for sound-hard  scatterers.  Denoting $\Omega$ the computational domain containing the array of scatterers, the real problem consists to solve in $\Omega$
\begin{subnumcases}{\label{AcousticD}}
\ds
\rho\,\frac{\partial {\bva}}{\partial t}=-{\bf \nabla}p,\label{Acoustic1D}\\
[6pt]
\ds
\chi\,\frac{\partial p}{\partial t}=-{\grad}.\bva,\label{Acoustic2D}\\[10pt]
\grad p\; . {\bf N}=0\;\textup{ on the boundaries of the scatterers,}
\end{subnumcases}
with ${\bva}=(v_x,v_y)^T$  the acoustic velocity and $p$  the acoustic pressure ($\chi$ is often written in terms of the sound speed  $c$: $\chi=(\rho c^2)^{-1}$). The Neumann boundary condition $\grad p\; . {\bf N}=0$, with $\bf N$ the vector locally normal to the scatterer boundaries, accounts  for a large mass density of the scatterers; it results a vanishing normal velocity at their boundaries.

\begin{figure}[h]
\centering
\includegraphics[width=.45\columnwidth]{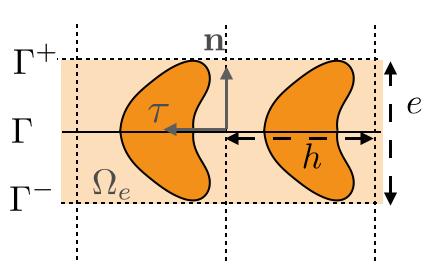} \hspace{1cm}
\includegraphics[width=.45\columnwidth]{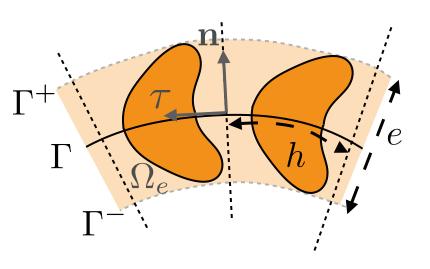} 

\small (a) \hspace{7cm} (b) 
\caption{ In the homogenized problem, the array of sound-hard scatterers  located periodically along a mean line $\Gamma$ is replaced by a domain $\Omega_e$. $\Omega_e$ is  delimited by the lines $\Gamma^-$ and $\Gamma^+$ across which the jump conditions (\ref{JCa}) apply. (a) in \cite{Marigo16b}, the jump conditions are expressed for a straight line $\Gamma$. (b) we extend these conditions to a curved line $\Gamma$; $\Gamma^-$ and $\Gamma^+$ are deduced from $\Gamma$ by a translation of $\pm e/2$ along the normal vector $\bf n$ to $\Gamma$; the  normal and tangent vectors to $\Gamma^\pm$ at the resulting points are the same than at the initial point on $\Gamma$. 
} 
\label{Fig2}
\end{figure}

If the central wavelength  (or the minimum one) $\lambda_0$ imposed by the wave source is much larger than $h$, the real problem can be replaced by an equivalent homogenized problem, owing to the introduction of the  small parameter 
\beq\label{defep}
\varepsilon\equiv 2\pi \frac{h}{\lambda_0}.
\eeq 
In \cite{Marigo16b}, such a homogenization has been proposed; the problem ends up with jump conditions for the pressure and for the normal velocity. In this reference, the derivation is performed for a line $\Gamma$ being straight (Fig. \ref{Fig2}(a)). Here, we heuristically extend this result to a curved line $\Gamma$, just by replacing locally the jump conditions expressed in Cartesian coordinates by jump conditions expressed in the local coordinates defined by the normal and tangent vectors to $\Gamma$ (Fig. \ref{Fig2}(b), and we shall discuss  the validity of this extension in this paper). 

Next, in \cite{Marigo16b} (see also the previous works of \cite{Delourme1,Delourme2}), the homogenized problem is shown to be associated with a satisfactory equation of energy conservation if the jump conditions are expressed across an ``enlarged" interface of interior $\Omega_e$ delimited by the two lines $\Gamma^-$ and $\Gamma^+$. Doing so allows us to define a positive interface energy, thus guaranties notably the well-posedness  in the time domain, which is necessary for our present purpose. The thickness of the interface is {\em a priori} arbitrary as soon as it guaranties  a positive interface energy. In this paper, following \cite{Marigo16b}, it is taken equal to the thickness $e$ of the array of Neumann scatterers.
We restrict ourselves to scatterers symmetrical {\em w.r.t.} the direction perpendicular to $\Gamma$. In this case, with $(\bn,{\bm \tau})$ denoting the local normal and tangent unitary vectors to $\Gamma$, the jump conditions read
\begin{equation}\left\{
\begin{array}{l}
\ds \jump{p}=
B \left< \dr{p}{n}\right>,\label{JC1}\\
[9pt]
\ds \jump{v_n}=
C_{1}\left<  \dr{v_n}{n}\right> +C_{2}\left<  \dr{v_\tau}{\tau}\right>,\end{array}\right.\label{JCa}
\end{equation}
where $v_n={\bva}.{\bf n}$, $v_\tau={\bva}.{\bm \tau}$,  and for any function $f$, $\partial_n f=\grad f.{\bf n}$, $\partial_\tau f=\grad f.{\bm \tau}$.   Also, we define 
\beq
\jump{p}={f}_{|\Gamma^+}-{f}_{|\Gamma^-}, \quad \textup{and}\; \left<   f\right> =\frac{1}{2}\left(f_{|\Gamma^-}+f_{|\Gamma^+}\right),
\eeq
being the jump of $f$ across the homogenized interface and the mean value of $f$ respectively  (the values $f_{|\Gamma^-}$ and $f_{|\Gamma^+}$ are defined locally  on  $\Gamma^-$ and $\Gamma^+$ along the $\bf n$ direction). In the above expressions, $B$, $C_1$  and $C_2$  are interface parameters which  depend on the shape of the scatterers only. As defined in (\ref{JCa}), these parameters have the dimension of lengths.
Finally,  the homogenized problem consists in solving in $\Omega\backslash \Omega_e$ 
\begin{subnumcases}{\label{Acoustic}}
\ds
\rho\,\frac{\partial {\bva}}{\partial t}=-{\bf \nabla}p,\label{Acoustic1}\\
[6pt]
\ds
\chi\,\frac{\partial p}{\partial t}=-{\grad. \bva},\label{Acoustic2}\\[6pt]
\textup{Jump conditions (\ref{JCa}) across $\Omega_e$.}
\end{subnumcases}
In the following section, the numerical resolution of the homogenized problem (\ref{Acoustic}) is specifically addressed. 


\section{Numerical methods}\label{SecNum}

The numerical scheme used to solved (\ref{Acoustic}) is implemented using the  first-order hyperbolic system
\begin{equation}
\frac{\partial}{\partial t}{\bU}+{\bf A}\frac{\partial}{\partial x}{\bU}+{\bf B}\frac{\partial}{\partial y}{\bU}={\bf 0},
\label{EDP}
\end{equation}
where (\ref{EDP}) is 
deduced from  (\ref{Acoustic}) by setting
\begin{equation}
{\bU}=
\left(
\begin{array}{l}
v_x\\
v_y\\
p
\end{array}
\right),
\hspace{0.3cm}
{\bf A}=
\left(
\begin{array}{ccc}
0 & 0 & 1/\rho\\
0 & 0 & 0\\
1/\chi & 0 & 0
\end{array}
\right),
\hspace{0.3cm}
{\bf B}=
\left(
\begin{array}{ccc}
0 & 0 & 0\\
0 & 0 & 1/\rho\\
0 & 1/\chi & 0
\end{array}
\right).
\label{UAB}
\end{equation}
The ADER-$r$ scheme is used \cite{Schwartzkopff04} to integrate (\ref{EDP}). It is an explicit and two-step finite-difference scheme of order $r$ in both space and time; here we use $r=4$, which amounts to a fourth-order Lax-Wendroff scheme \cite{Lorcher05}. It is dispersive of order 4 and dissipative of order 6 \cite{Strickwerda99}. Finally, it is stable under the CFL condition $\beta=c\frac{\Delta t}{\Delta x}\leq 1$ (in two dimensions). 

The solution  $\bU$ is discretized in space on a uniform Cartesian grid with mesh sizes $\Delta x=\Delta y$, and in time with a time step $\Delta t$, and we denote  $\bU_{i,j}^n$  the discretized numerical value of $\bU$  at $M(i,j)$ at $(x=i\Delta x,y=j\Delta y)$ and at time $t=n\Delta t$. The calculation of $\bU_{i,j}^n$ involves  a  stencil of $(r+1)^2$  nodes centered at $M$  (25 nodes for ADER 4), which is written formally as the time-marching   
\begin{equation}
{\bU}^{n+1}_{i,j}={\cal H}\left({\bU}^n_{i_0,j_0}\right),\hspace{0.6cm} \left\{i_0-i,j_0-j\right\}\in\left\{-2,\cdots,+2\right\},
\label{ADER 4}
\end{equation}
with ${\cal H}$ being deduced from (\ref{EDP}) and (\ref{UAB}).

\vspace{.3cm}
The jump conditions (\ref{JCa}) of the homogenized problem are discretized by the scheme (\ref{ADER 4}). We introduce now a numerical method that incorporates these conditions in the numerical scheme, for a negligible additional cost. 


\subsection{Modified ADER scheme in the neighborhoods of $\Omega_e$}\label{SecNumScheme}

\begin{figure}[h!]
\centering
\includegraphics[width=.495\columnwidth]{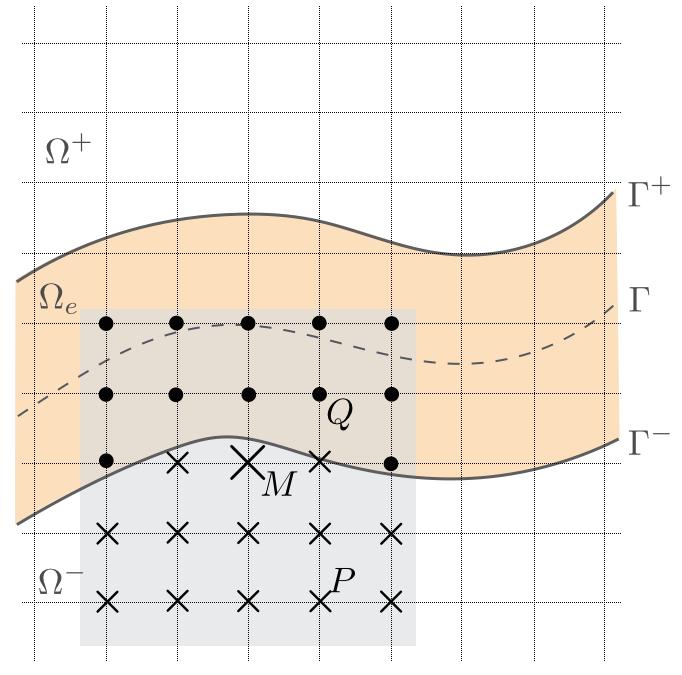} 
\includegraphics[width=.49\columnwidth]{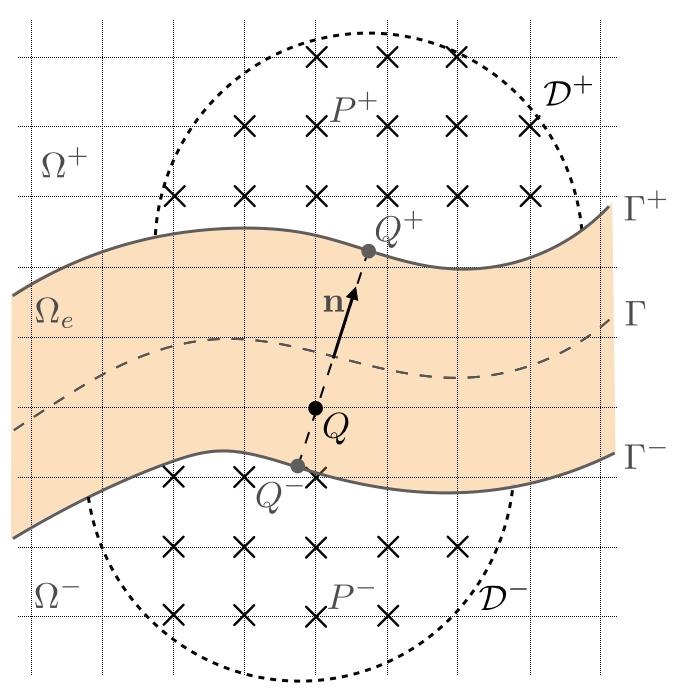} 

\small (a) \hspace{5.5cm} (b)
\caption{\label{Fig4} (a) Example of irregular point $M$; in the stencil of $M$ (grey region),  $Q$ ($\bullet$) are in $\Omega_e$ (the usual nodes, in $\Omega^\pm$ are denoted $P$ ($\times$)). (b) Construction of the nodes $\Pn^\pm$ ($\times$) in $\Omega^\pm$ used to  calculated the modified value $\bU^*(Q)$. With $Q^\pm$  the  projections of $Q$ on $\Gamma^\pm$ along the normal $\bf n$, the $\Pn^\pm$ are the nodes in the half-disks ${\cal D}^\pm$ of centers $Q^\pm$ ($\Pn^\pm$ belong to  $\Omega^-$ and $\Omega^+$).
} 
\end{figure}

To solve the homogenized problem (\ref{Acoustic}), the scheme (\ref{ADER 4}) has to be adapted.
Because no constitutive law is defined in  the thick interface  $\Omega_e$,  the solution $\bU$ in this region  is not defined.  
Thus, we shall first distinguish so-called {\em regular} and {\em irregular} points $M$ depending on the position of the   stencil of $M$  with respect to the  interface.   
In the following, we denote $\Omega^\pm$ the subdomains of $\Omega$ above $\Gamma^+$ and below $\Gamma^-$ (Fig. \ref{Fig4}), and $\Omega^+\cup \Omega^-=\Omega\backslash \Omega_e$. 
\vspace{.2cm}
  
\noindent When all the nodes $\Pn$ of the stencil at $M$ fall in the same physical medium ($\Omega^-$ or $\Omega^+$), the point $M$ is called   a {\it regular point}. For these points,  (\ref{ADER 4}) can be used straightforwardly using  $\bU(M)=\bU_{i,j}^n$ and $\bU(\Pn)=\bU_{{i_0},{j_0}}^n$ being  the discretized values  of the continuous solution $\bU$; we call them  {\em direct values}. 

\vspace{.2cm}
\noindent
In the  neighborhood of $\Omega_e$, it  happens that  the stencil of $M$ crosses $\Gamma^-$ or $\Gamma^+$ (Fig. \ref{Fig4}(a)). 
 It results that  nodes $Q$ in the stencil of $M$ belong to $\Omega_e$. Such points $M$ are called {\em irregular} since  $\bU(M)=\bU_{i,j}^n$  in (\ref{ADER 4}) requires the values of $\bU$ at the nodes $Q$,  and $\bU$ is not defined in $\Omega_e$. In the forthcoming discussion, we focus on an irregular node $Q$ close to $\Gamma^-$: it means that $Q$ is used in the stencils of grid nodes in $\Omega^-$.

\vspace{.3cm}
 
To account for the regular and irregular points, it is possible to adapt (\ref{ADER 4}) in a simple way. This is done   by attributing  {\em modified values} $\bU^*(Q)$  for the  points $Q$   in  $\Omega_e$, and the direct values otherwise. Specifically, the modified ADER 4 is modified as 
\begin{equation}
\left\{\begin{array}{l}
 {\bU}^{n+1}_{i,j}={\cal H}\left(\tilde {\bU}^n_{i_0,j_0}\right),\hspace{0.6cm} \left\{i_0-i,j_0-j\right\}\in \left\{-2,\dots,+2\right\}, \\[10pt]
 \tilde {\bU}^n_{i_0,j_0}= {\bU}^n_{i_0,j_0}, \quad (x_{i_0},y_{j_0})\notin \Omega_e,\\[10pt]
\tilde {\bU}^n_{i_0,j_0} ={{\bU}^{n*}_{i_0,j_0}}, \quad (x_{i_0},y_{j_0}) \in \Omega_e.
\end{array}\right.
\label{ADER 4-mod}
\end{equation}

In (\ref{ADER 4-mod}), the numerical values of the solution are affected to all the nodes in the computational domain (with  direct values $\bU$ in $\Omega^-$ and $\Omega^+$ and modified values $\bU^*$ in $\Omega_e$). 
We shall see in the following section that   the modified values (at the points $Q$) are expressed in terms of the direct values at  points $\Pn^\pm$  in $\Omega^\pm$ (Fig. \ref{Fig4}(b), the choice of the $\Pn^\pm$ is incidental at this stage).  It follows that  (\ref{ADER 4-mod}) is solved implicitly on the direct values only:
  the solution in $\Omega_e$ is not questioned, as expected ($Q$ does not appear as the center of a stencil in (\ref{ADER 4-mod})). 


\subsection{Construction of the modified values $\bU^*$ at the nodes $Q$ in  $\Omega_e$ }\label{SecNumESIM}

As previously said, the solution  in $\Omega_e$ is undefined   ($\bU^*(Q)$ does not exist). Only the jump conditions across $\Omega_e$ make sense. Thus, we shall start by defining $\bU^*(Q)$ and this will be done using $Q^\pm$ being the projections of $Q$ on $\Gamma^\pm$ along $\bn$, and using the jump conditions (\ref{JCa}) which apply between $Q^+$ and $Q^-$. 
Next, because $Q^\pm$ are not nodes (except by casuality),  the solutions at $Q^\pm$ will be expressed in terms  of two sets of direct values $\bU(\Pn^\pm)$. This is the  meaning of the nodes $\Pn^\pm$, chosen in the vicinity  of $Q^\pm$. 

\vspace{.3cm}
The construction of $\bU^*(Q)$ presented in the forthcoming  section applies for any sets of $\Pn^\pm$ being nodes of $\Omega^\pm$ in the vicinities of $Q^\pm$. 
 Fig. \ref{Fig4}(b) illustrates our choice:  $\Pn^\pm$ are the nodes of $\Omega^\pm$  in the disks ${\cal D}^\pm$ centered at $Q^\pm$ and of radius $d$. In practice, $d\simeq 3.5 \Delta x$  and this is discussed further in Sec. \ref{sec33}.


\subsubsection{The ingredients of the construction of $\bU^*(Q)$}

In this section, the construction of $\bU^*(Q)$ is presented formally. It will be detailed precisely in the Sections \ref{step1} to \ref{step6}. Let us recall that $Q$ is used for time-marching at points of $\Omega^-$, and thus is close to $\Gamma^-$.

\vspace{.3cm} 
Because the solution is not defined in $\Omega_e$, we start  by defining  $\bU^*(Q)$ as  the smooth extension of the solution in $\Omega^-$. This is done using a Taylor expansion for $Q$  in the neighborhood of $Q^-$, written formally
\beq\label{to1}
\bU^*(Q)=\mT(Q,Q^-) \bW(Q^-).
\eeq
In one dimension at order 1, $\bU^*(Q)=\bU(Q^-)+(x_Q-x_{Q^-})\partial_x \bU(Q^-)$;   $\bW(Q^-)$ encapsulates the weights $(\bU(Q^-),\partial_x \bU(Q^-) )$ while $\mT(Q,Q^-)$ are the polynomial forms depending on both $Q$ and $Q^-$, here $1$ and $(x_Q-x_{Q^-})$. 
It is worth noting  that (\ref{to1}) introduces a disymmetry between $\Omega^-$ and $\Omega^+$. Indeed,  if $\bU^*(Q)$ is a smooth extension of the solution in $\Omega^-$, it cannot be a smooth extension of the solution in $\Omega^+$ because of  the jump conditions.  
These jump conditions  apply between $Q^-$ and $Q^+$ but $Q^\pm$ do not coincide with nodes. Thus, we use the sets of nodes $\Pn^\pm$ in the vicinities of $Q^\pm$, and the Taylor expansions 
\beq\label{to2}
\bU(\Pn^\pm)=\mT(\Pn^\pm,Q^\pm) \bW(Q^\pm),
\eeq
which involves the direct values $\bU(\Pn^\pm)$. 

From  (\ref{to1}-\ref{to2}), it is visible that $\bU^*(Q)$ can be expressed as a function of $\bU(\Pn^\pm)$ if a relation between  $\bW(Q^+)$ and $\bW(Q^-)$ is established. This relation will obviously involve the jump conditions. 
Before doing so, the number of terms in $\bW$ is reduced owing to high-order compatibility conditions, deduced from the initial condition $\nabla \wedge {\bf v}={\bf 0}$ (coming from (\ref{Acoustic1}) and valid in $\Omega^\pm$). The  new vector $\bV$ collects the reduced numbers of unknowns of $\bW$, with 
\beq\label{to3}
\bW(Q^\pm)=\mG \bV(Q^\pm), \quad \bW(\Pn^\pm)=\mG \bV(\Pn^\pm),
\eeq   
and $\mG$ is a constant matrix. Finally, the  jump conditions are used to get a relation between $\bV(Q^+)$ and $\bV(Q^-)$, written  
\beq \label{to5}
\bV(Q^+)=\fu\left[\bV(Q^-)\right].
\eeq

The relations (\ref{to1}) to (\ref{to5}) allow to conclude. First, $\bU(\Pn^\pm)$ can be expressed as a function of $\bV(Q^-)$ only. From (\ref{to2}), we have $\bU(\Pn^-)=\mT(\Pn^-,Q^-)\mG \bV(Q^-)$. For $\bU(\Pn^+)$, it starts the same, with $\bU(\Pn^+)=\mT(\Pn^+,Q^+)\mG \bV(Q^+) $  and  (\ref{to5}) allows to conclude (and (\ref{to5}) is essential since it encapsulates the jump conditions). Using these relations for all the nodes $\Pn^\pm\in {\cal D}^\pm$ and collecting the direct values $\bU(\Pn^\pm)$ in a single vector ${\bf U}=\left(\bU(\Pn^-)_{|\Pn^-\in {\cal D}^-},\bU(\Pn^+)_{|\Pn^+\in {\cal D}^+}\right)$, the formal relation ${\bf U}={\bf M} \bV(Q^-)$ can be inverted to get 
 \beq\label{to4}
\bV(Q^-)= {\bf M}^{-1}
 \left(\begin{array}{c}
 \bU(\Pn^-)\\ \bU(\Pn^+)
 \end{array}\right). 
 \eeq
Finally, (\ref{to4}) is injected in (\ref{to1}), using (\ref{to3}), to get the modified values $\bU^*(Q)$ as a function of the direct values $\bU(\Pn^\pm)$
\beq\label{toend}
\bU^*(Q)=\mT(Q,Q^-)\; \mG \; {\bf M}^{-1} \left(\begin{array}{c}
 \bU(\Pn^-)\\ \bU(\Pn^+)
 \end{array}\right).
\eeq
We shall now detail the  steps in the construction of $\bU^*(Q)$ as implemented numerically for   $k$-th order Taylor expansions.  

 
\subsubsection{The Taylor expansions, Eqs. (\ref{to1})-(\ref{to2})} \label{step1}

Here, we simply specify the notations in (\ref{to1}) and (\ref{to2}) to get Taylor expansions at the order $k$.   The matrix $\mT_k$   of  $k$-th order  expansions  for $Q$ near  $Q^-$ reads 
\begin{equation}  
\mT_k(Q,Q^-)=\left({\bf I}_3,\cdots,\frac{1}{\ell !\,(\ell-m)!}\left(x_Q-x_{Q^-}\right)^{\ell-m}\left(y_Q-y_{Q^-}\right)^m{\bf I}_3,\cdots,\frac{\left(y_Q-y_{Q^-}\right)^k}{k !}{\bf I}_3\right),
\label{MatTaylor}
\end{equation}
with   $\ell=0,\cdots,k$ and $m=0,\cdots,\ell$ and  ${\bf I}_3$  the $3 \times 3$ identity matrix. With $(\ell +1)$ polynomial forms at  the  order $\ell$, $\mT_k$ is a    $3\times \nT$ matrix, with $\nT=3(k+1)(k+2)/2$.
Next, we collect in a single vector $\bW_k$ the $\nT$ limit values of ${\bU}$ 
and of its successive spatial derivatives up to the $k$-th order, at $Q^\pm$:
\begin{equation}
\bW_k(Q^\pm)=\lim_{Q^\pm \in\Omega^\pm}
\left(
{\bU}^T,
...,\,
\frac{\partial^\ell}{\partial x^{\ell-m}\,\partial y^m}\,{\bU}^T,
...,\,
\frac{\partial^k}{\partial y^k}\,{\bU}^T
\right)^T,
\label{Uk}
\end{equation}
with $\ell=0,\,...,\,k$ and $m=0,\,...,\,\ell$. 
The modified value  $\bU^*(Q)$  is defined as a smooth extension of the solution in $\Omega^-$
\begin{equation}
\ds {\bU}^*(Q) = \ds \mT_k(Q,Q^-)\,\bW_k(Q^-),
\label{SolMod}
\end{equation}
and   $\bU^*(Q)$ appears to depend on $k$. As previously said, once this definition is chosen, $\bU^*(Q)$ cannot be a smooth extension of the solution in $\Omega^+$ because of  the jump conditions. 

\vspace{.3cm}

The story is different for the Taylor extensions of the direct values $\bU(\Pn^\pm)$ written crudely in (\ref{to2}). Because $\bU(\Pn^\pm)$ are the discretized versions of the exact solution, their Taylor expansions have to be written as  approximations, namely 
\beq\label{ti2}
\bU(\Pn^\pm)=\mT_k(P^\pm,Q^\pm)\bW_k(Q^\pm)+\grandO.
\eeq 


\subsubsection{High-order compatibility condition, Eq. (\ref{to3})}  \label{step2}

The equation (\ref{Acoustic1}) provides a compatibility condition, $\nabla \wedge {\bf v}={\bf 0}$, which tell us  that the fluid is irrotational in $\Omega^\pm$. 
Assuming  sufficiently smooth solutions  in $\Omega^\pm$, this relation can be differentiated $(\ell-1)$ times {\em w.r.t.} to $x$ and $y$:
\begin{equation}
\frac{\partial^\ell v_y}{\partial x^{\ell-m-1}\partial y^{m+1}}=\frac{\partial^\ell v_x}{\partial x^{\ell-m}\partial y^{m}},\hspace{0.6cm}\ell\geq 1,\hspace{0.1cm}m=0,\cdots,\ell-1,
\label{Compatk}
\end{equation} 
whose version with $\ell=1,m=0$ is the originate condition. 
Doing so for  $\ell=1,\dots, k$  provides $k(k+1)/2$ high-order compatibility conditions. The equations  are valid everywhere outside $\Omega_e$, and in particular at $Q^\pm$. 
This allows to use vectors $\bV_k$ containing  only the remaining independent  derivatives, thus being reduced in size (to $\nC=(k+1)(k+3)$ components). The $\bV_k$ are linked to $\bW_k$ by  
\begin{equation}
\bW_k(Q^\pm)={\bf G}_k\,\bV_k(Q^\pm),
\label{Compatibilite}
\end{equation}
where ${\bf G}_k$ is a $\nT \times \nC$ matrix (an algorithm to compute ${\bf G}_k$ can be found in  \cite{Lombard04}, see the appendix A in this reference). 
Now,  (\ref{ti2}) can be written  as a function of $\bV_k(Q^\pm)$, using (\ref{Compatibilite}), 
\beq\label{ti3}
\bU(\Pn^\pm)=\mT_k(\Pn^\pm,Q^\pm) {\bf G}_k 
\bV_k(Q^\pm)
+\grandO.
\eeq 


\subsubsection{High-order order jump conditions, (\ref{to5})} \label{step3}

\begin{figure}[htbp]
\centering
\includegraphics[width=.7\columnwidth]{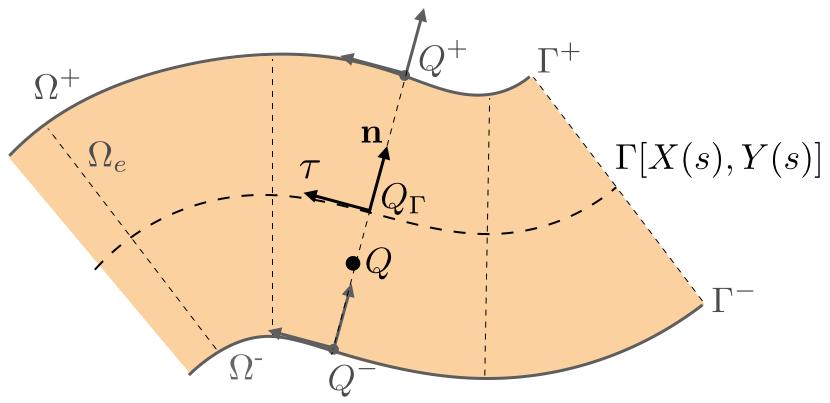}
\caption{\label{Figtheta} 
Parametrization of $\Gamma$ with the parameter $s$; the  vectors ($\bn,{\bm \tau}$) locally normal and tangent to $\Gamma$ at $\Qm[X(s), Y(s)]$ are also normal and tangent to $\Gamma^\pm$ at $Q^\pm$ by construction.
}
\end{figure}

To express the jump conditions between $Q^-$ and $Q^+$, it is useful to come back to the construction of $\Gamma^\pm$ and of  $Q^\pm$ (Fig. \ref{Figtheta}). Because $\Gamma^\pm$ are deduced from $\Gamma$ by the translations of $\pm e/2$ along $\bn$,  the   vectors locally normal and tangent to $\Gamma^\pm$ at $Q^\pm$ are the same, and the same as the vectors $(\bn,{\bm \tau})$, locally normal and tangent  to $\Gamma$ at $\Qm$. 
In practice, we use a parametrization of $\Gamma$ with $s$, with $[(X(s),Y(s)]$ the coordinates of $\Qm$. 
  We denote $\bn=(n_1,n_2)$ and ${\bm{\tau}}=(t_1,t_2)$, whence $n_1=t_2=Y'(s)$, $t_1=-n_2=-X'(s)$, with prime denoting the derivative.

The jump conditions (\ref{JCa}) can be encapsulated in the matrix form
\begin{equation}
{\mE}_0 \;\left[ {\bW}_0(Q^+)-{\bW}_0(Q^-)\right]=\frac{1}{2}{\mF}_1\,\left[{\bW}_1(Q^-)+{\bW}_1(Q^+)\right],
\label{JC0}
\end{equation}
where the matrices  $\mE_0$ and $\mF_1$ depend  on the geometry of the interface because of their dependance on  $(\bn, {\bm{\tau}})$. 
  $\mE_0$ is a $2\times 3$ matrix defined by 
\begin{equation}
{\bf E}_0=\left(  
\begin{array}{ccc}
n_1 & n_2 & 0\\
0 & 0 & 1
\end{array}
\right),
\label{E00}
\end{equation}
and $\mF_1$  is a $2 \times 9$ matrix whose non-zero components are
\begin{equation}
\begin{array}{l}
\ds \mF_1(1,4)=C_{1}n_1^2+C_{2}n_2^2,\\
[8pt]
\ds \mF_1(1,5)=\mF_1(1,7)=C_{1}n_1n_2+C_{2}t_1t_2,\\
[8pt]
\ds \mF_1(1,8)=C_{1}n_2^2+C_{2}t_2^2,\\
[8pt]
\ds \mF_1(2,6)=Bn_1,\\
[8pt]
\ds \mF_1(2,9)=Bn_2.
\end{array}
\label{F11}
\end{equation}
Now, the goal is to obtain an extended version of  the jump conditions involving derivatives of $\bU$ up the $k$-th order, namely involving $\bW_k$. For this purpose, (\ref{JC0}) is differentiated {\em w.r.t} the time $t$ and the parameter $s$, and using 
\beq\left\{
\begin{array}{l}
\ds \dr{}{t}\bW_\ell=-{\bf A}\dr{}{x}\bW_\ell--{\bf B}\dr{}{y}\bW_\ell,\\
[8pt]
\ds  \dr{}{s}=X'(s)\dr{}{x}+Y'(s)\dr{}{y},
\end{array}\right.
\eeq
where the first relation holds for any $\ell$. Applying to (\ref{JC0}) the chain-rule 
\beq
\frac{\partial^{(\ell-i)}}{\partial t^{(\ell-i)}} \frac{\partial^i}{\partial s^i}\left[{\mE}_0 \left({\bW}_0(Q^+)-{\bW}_0(Q^-)\right)-\frac{1}{2} \mF_1\,\left({\bW}_1(Q^+)+{\bW}_1(Q^-)\right)\right]={\bf 0}, \quad i=0,\dots, \ell,
\eeq 
provides $(\ell+1)$ matrix relations, or $2(\ell +1)$ scalar relations, at each order $\ell$. These relations involve spatial derivatives up to  the order $(\ell +1)$ and the chain-rule is stopped at the order $k$. The  $\nJ=(k+1)(k+2)$ obtained relations for $\ell=0$ to $k$ are collected in the matrix relation 
 
\begin{equation}
\mE_k\,\left[{\bW}_k(Q^+)-{\bW}_k(Q^-)\right]=\frac{1}{2}\mF_{k+1}\,\left[{\bW}_{k+1}(Q^+)+{\bW}_{k+1}(Q^-)\right].
\label{JCk}
\end{equation}
The matrices $\mE_k$ and $\mF_{k+1}$ are non trivial, and their  computation  has been automatized thanks to computer algebra tools.
It is worth noting that they depend on the first $k$-th derivatives of $X(s)$ and $Y(s)$, thus they encapsulate informations on the interface shape well beyond its position only.

\vspace{.2cm}

In (\ref{JCk}),  $\bW_{k+1}$ involves  spatial derivatives up to the order $(k+1)$, and this order is not considered in the resolution at the order $k$.  Considering  $\overline{\mF_{k+1}}$ the restriction of $\mF_{k+1}$ obtained by removing the last $(\nT+1)$-th column in the matrix $\mF_{k+1}$, (\ref{JCk}) simplifies to 
\begin{equation}
{{\mE}^+_{k}}\,{\bW}_k(Q^+)={{\mE}^-_{k}}\,{\bW}_k(Q^-),
\label{JCfinal}
\end{equation}
where  $\mE_k^\pm\equiv \mE_k\mp \frac{1}{2}\overline{\mF_{k+1}} $ are two $\nJ\times \nT$ matrices.
Combining  (\ref{Compatibilite}) and  (\ref{JCfinal}) yields 
\begin{equation}
{\bf S}^+_k\,{\bV}_k(Q^+)={\bf S}_k^-\,{\bV}_k(Q^-),
\label{SV}
\end{equation}
with  the $\nJ\times \nC$ matrices ${\bf S}_k^\pm={{\mE}_k^\pm}\,{\bf G}_k$ being smaller than the $\nJ\times \nT$ $\mE_k^\pm$. With $\nJ=(k+1)(k+2)$ and $\nC=(k+1)(k+3)$,  the system (\ref{SV}) is underdetermined. It is inverted in the sense of the least-squares sense  using Singular Value Decomposition, leading to
\begin{equation}
{\bV}_k(Q^+)=\left(\left({\bf S}_k^+\right)^{-1}\,{\bf S}_k^-\,|\,{\bf K}_{{\bf S}_k^+}\right)
\left(
\begin{array}{c}
\displaystyle
{\bV}_k(Q^-)\\
[8pt]
\displaystyle
{\bf \Lambda}_k
\end{array}
\right),
\label{SVD}
\end{equation}
where $({\bf S}_k^+)^{-1}$ is the least-square pseudo-inverse of ${\bf S}_k^+$, 
${\bf K}_{{\bf S}_k^+}$ is the matrix filled with the kernel of ${\bf S}_k^+$, and ${\bf \Lambda}_k$ is a set of the $(\nC-\nJ)$ Lagrange multipliers which are the coordinates of ${\bV}_k(Q^+)$ onto the kernel. A singular value decomposition of ${\bf S}_k^+$ is used to build $({\bf S}_k^+)^{-1}$ and the kernel ${\bf K}_{{\bf S}_k^+}$ \cite{Press92}.\\


\subsubsection{Final step in the construction of modified values, (\ref{to4})-(\ref{toend})} \label{step6}

We want to express $\bV_k(Q^-)$ as a function of the set of direct values $\bU(\Pn^\pm)$ (we denote $n^\pm$ the numbers of nodes $\Pn^\pm$). For the set of nodes $\Pn^-$ in $\Omega^-$,   (\ref{ti2}) is simply re-written 
\beq\label{Taylor1}
\bU(\Pn^-)=\mT_k(\Pn^-,Q^-) {\bf G}_k \,\left({\bf 1}\,|\,{\bf 0}\right)
\left(
\begin{array}{c}
\bV_k(Q^-)\\
[8pt]
{\bf \Lambda}_k
\end{array}
\right)
+\grandO.
\eeq 
In the above relation, ${\bf 1}$ stands for  the $\nC \times \nC$ identity matrix and ${\bf 0}$ for the $\nC\times(\nC-\nJ)$ zero matrix. For the set of nodes $\Pn^+$ in $\Omega^+$, we use (\ref{SVD}) in  (\ref{ti3}), whence  
\beq
\bU(\Pn^+)=\mT_k(\Pn^+,Q^+) {\bf G}_k \,\left[\left({\bf S}_k^+\right)^{-1}\,
{\bf S}_k^-\,|\,{\bf K}_{{\bf S}_k^+}\right]
\left(
\begin{array}{c}
\displaystyle
{\bV}_k(Q^-)\\
[8pt]
\displaystyle
{\bf \Lambda}_k
\end{array}
\right)
+\grandO.
\label{Taylor2}
\end{equation}
Now, we collect in a single vector $\bf U$ the $(n^-+n^+)$ vectors $\bU(\Pn^\pm)$
\beq
{\bf U}=\left(\begin{array}{l}
\bU(\Pn_1^-)\\
\vdots \\
\bU(\Pn_{n^-}^-)\\[8pt]
\bU(\Pn_1^+)\\
\vdots \\
\bU(\Pn_{n^+}^+)
\end{array}\right),
\eeq
and use (\ref{Taylor1}) and (\ref{Taylor2}) written in the matrix form 
\begin{equation}
{\bf U}={\bf M}
\left(
\begin{array}{c}
{\bV}_k(Q^-)\\
[8pt]
{\bf \Lambda}_k
\end{array}
\right)
+\grandO,
\label{Taylor3}
\end{equation}
where 
\beq
{\bf M}\equiv
\left(
\begin{array}{c}
\mT_k(\Pn^-_1,Q^-) {\bf G}_k \,\left({\bf 1}\,|\,{\bf 0}\right)\\
\vdots \\
\mT_k(\Pn^-_{n^-},Q^-) {\bf G}_k \,\left({\bf 1}\,|\,{\bf 0}\right)\\[8pt]
\mT_k(\Pn^+_1,Q^+) {\bf G}_k \,\left[\left({\bf S}_k^+\right)^{-1}\,{\bf S}_k^-\,|\,{\bf K}_{{\bf S}_k^+}\right]\\
\vdots\\
\mT_k(\Pn^+_{n+},Q^+) {\bf G}_k \,\left[\left({\bf S}_k^+\right)^{-1}\,{\bf S}_k^-\,|\,{\bf K}_{{\bf S}_k^+}\right]
\end{array}
\right)
\label{defM}
\eeq
is a $(3 (n^-+ n^+) \times (2\,\nC-\nJ)$ matrix. To ensure that the system (\ref{Taylor3}) is overdetermined, the radius $R_{\cal D}$ of the discs ${\cal D}^\pm$ has to be chosen in order to ensure that   
\begin{equation}
\gamma(R_{\cal D},k)\equiv\frac{\textstyle 3\,(n^-+n^+)}{\textstyle (k+1)(k+4)}\geq 1,
\label{RayonPatate}
\end{equation}
and this condition will be further discussed in Section \ref{sec33}.
The condition (\ref{RayonPatate}) being ensured,  (\ref{Taylor3}) is solved using the least-squares inverse ${\bf M}^{-1}$ of ${\bf M}$. The  Lagrange multipliers ${\bf \Lambda}_k$ have been  incorporated in the construction of ${\bf M}$, but they are not needed to build the modified value. Thus, they are removed  using the $(\nT-\nC)\times 3\,(n^-+n^+)$ restriction $\overline{{\bf M}^{-1}}$ of ${\bf M}^{-1}$, to get only 
$$\bV_r(Q^-)=\overline{{\bf M}^{-1}} {\bf U},$$ and finally 
using  (\ref{SolMod})
\begin{equation}
{\bU}^*(Q)={\mT}_k(Q,Q^-)\,{\bf G}_k \,\overline{{\bf M}^{-1}}\,{\bf U}.
\label{UIJ*}
\end{equation}
The matrices $\mT_k(\Pn^\pm,Q^\pm)$, $\mG_k$ and $\left[\left({\bf S}_k^+\right)^{-1}\,{\bf S}_k^-\,|\,{\bf K}_{{\bf S}_k^+}\right]$ are involved in 
(\ref{UIJ*}) through $\overline{{\bf M}^{-1}}$.


\subsection{ Comments and practical details}\label{sec33}

\begin{enumerate}
\item The described algorithm  is applied to the irregular points in $\Omega^\pm$. 
The sizes of the matrices involved are summarized in table \ref{TabMatrices}. Since the jump conditions do not vary with time, the evaluation of the matrices in (\ref{UIJ*}) is done during a preprocessing step. Only small matrix-vector products are therefore required at each time step. After optimization of the computer codes, this additional cost is made negligible, lower than 1\% of the time-marching of the ADER 4 scheme.

\item The matrix ${\bf M}$ in (\ref{Taylor3}) depends on the subcell positions of $Q^\pm$ inside the mesh and on the jump conditions expressed between $Q^+$ and $Q^-$, involving the local geometry and the curvature of $\Gamma^\pm$ at $Q^\pm$. Consequently, all these insights are incorporated in the modified value (\ref{UIJ*}), and thus also in the scheme.

\item The optimal order $k$ depends on the order $r$ of the scheme and on the jump conditions. Let us begin with the classical case of acoustics where the jump conditions do not involve spatial derivatives: for instance, $\jump{v_n}=0$ and $\jump{p}=0$. In this case, taking $k=r$ maintains a $r$-th order global accuracy \cite{Chiavassa11} (the criterion $k=r-1$ is even sufficient \cite{Gustafsson75}). In the non-classical case studied here, the jump conditions involve first-order spatial derivatives. After successive derivations, the higher-order terms are canceled in (\ref{JCfinal}), which introduces a loss of accuracy. To maintain the $r$-th order convergence, the order of the immersed interface method must be increased of one unity: $k=r+1\equiv 5$ for the ADER 4 scheme.

\item \label{EstimeVarEps} The simulations indicate that overestimation of $\gamma$ in (\ref{RayonPatate}) ensures the stability of the immersed interface method. Numerical experiments have shown that $d=3.5\,\Delta x$ is a good choice. Typically, it gives $n^\pm \approx 10$ and $\gamma \approx 4$. 
\end{enumerate}

\begin{table}[htbp]
\begin{center}
\begin{tabular}{|l|l|}  
\hline
$\nT=3\,(k+1)\,(k+2)/2$ \\
$\nC=(k+1)\,(k+3)<\nT$\\
$\nJ=(k+1)\,(k+2)<\nC$ \\
\hline
\hline
\end{tabular}
\\
\begin{tabular}{|l|l|}  
\hline
Quantity & Size\\
\hline
\hline
${\mT}_k $& $3 \times \nT$\\
${\bf G}_k $  & $\nT \times \nC$ \\
${\bf S}_k^\pm$ & $\nJ \times \nC$ \\
${\bf M}$     & $3\,(n^++n^-) \times (2\,\nT-2\,\nC-\nJ)$\\
$\overline{{\bf M}^{-1}}$ & $3\,(n^++n^-)\times (\nT-\nC)$\\
\hline
\end{tabular}
\caption{Quantities involved in the computation of the modified values (section \ref{SecNumESIM}).}
\label{TabMatrices}
\end{center}
\end{table}


\section{Numerical experiments}\label{SecExp}

In this section, we first validate the  immersed interface method implemented on the thick  interface as presented in the preceding section. This is done for  a plane wave at normal and oblique incidences on the interface. These cases allow for analytical solutions of the homogenized problem, already validated in \cite{Marigo16a,Marigo16d} in the harmonic regime. The extension to the time domain being done by means of discrete inverse Fourier transforms, we do not repeat the validation by comparison of the solutions of the homogenized and of the direct problems. We inspect the accuracy of the numerical scheme varying the order $k$ in the scheme, and discuss the convergence obtained with the estimated optimal value $k=5$.

Next, we exemplify the capability  of the homogenized problem  (\ref{Acoustic}) to mimic the real problem  (\ref{AcousticD}) in the time domain. The solutions of the direct and the homogenized problems are compared in the case of a source emitting a short pulse with a central frequency that we vary to inspect the robustness of the homogenized solution {\em w.r.t.} the small parameter $\varepsilon$, defined in (\ref{defep}) (and $\lambda_0$ will be defined latter). This is done for a straight and curved line array.

\vspace{.5cm}

The following characteristics hold for all our simulations:
\begin{itemize}
\item{  We consider arrays of typically 50 rectangular sound-hard scatterers in water. The periodicity of the array is $h=20$ m with a filling ratio of the scatterers $\varphi=0.5$; the thickness of the array is $e=20$ m. For these dimensions of the scatterers, the  interface  coefficients entering in the jump conditions  are 
\beq\label{param}
B=44.412, \quad C_{1}=10, \quad C_{2}=8.338. 
\eeq
(see \ref{SecIntPar}).
For water, we use $\rho=1000\,\mbox{kg/m}^3$ and $c=1500$ m/s.}
\item{
The numerical results have been performed in a   domain of extension 1200 m$\times$ 1200 m, discretized on $N_x \times N_y$ points; in practice, we used $N_x=N_y$. The time step follows from the CFL condition: $\Delta t=\beta \Delta x/c$, and we used $\beta=0.95$.  If not specified, the order  in the ADER 4 scheme is $k=5$. Most of the simulations have been performed with  $N_x=600$ ($\Delta x=\Delta y=2$ m) and, from the CFL condition, $\Delta t=1.27.10^{-3}$ s.}
\item{
Discrepancies between two solutions yielding the pressure fields $p_1$ and $p_2$ are given by the relative difference  $||p_1-p_2||/||p_1||$ (and $||.||$ refers to the discrete L$^2$- norm)}
\item{
The temporal signals are built using $h(t)$ being a combination of truncated sinusoids
\begin{equation}
h(t)=
\left\{
\begin{array}{l}
\ds \sum_{m=1}^4 a_m\,\sin(\beta_m\,\omega_0\,t)\quad \mbox{ if  }\, 0<t<\frac{1}{f_0},\\
[8pt]
0 \,\mbox{ otherwise}, 
\end{array}
\right.
\label{JKPS}
\end{equation}
where $\beta_m=2^{m-1}$, $\omega_0=2\pi\,f_0$; the coefficients $a_m$ are: $a_1=1$, $a_2=-21/32$, $a_3=63/768$, $a_4=-1/512$, ensuring $C^6$ smoothness of $h(t)$. The Fourier transform of (\ref{JKPS}) reads
\begin{equation}
{\hat h}(\omega)=\frac{\omega_0}{2\pi}\sum_{m=1}^4 a_m\beta_m \frac{e^{2i\pi \omega_0/\omega}-1}{\omega^2-\beta_m\,\omega_0^2},
\label{Fourier}
\end{equation}
with a maximum slightly greater than $f_0$ and a cut-off frequency $f_m$ at around 5$f_0$. $h(t)$ and $\hat h(\omega)$ are shown  in Figs. \ref{FigSource} for $f_0=10$ Hz. 

In the simulations, various central frequencies are considered: 
$f_0=2.5$, 5 and 10 Hz. The resulting values of $\varepsilon$ calculated using $k_0=2\pi f_0/c$ are  $\varepsilon=0.21$, 0.41 and 0.83. However the spectral content of the source being large, the wave packet contains wavenumbers  5 times smaller than $k_0$; these small scales are associated to $\varepsilon$ larger than 1 (up to 4). 

\begin{figure}[htbp]
\centering
\includegraphics[width=.4\columnwidth]{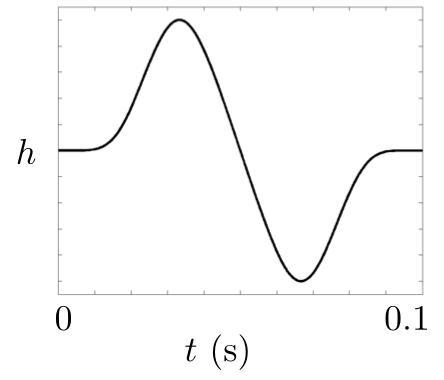} \hspace{.3cm}
\includegraphics[width=.4\columnwidth]{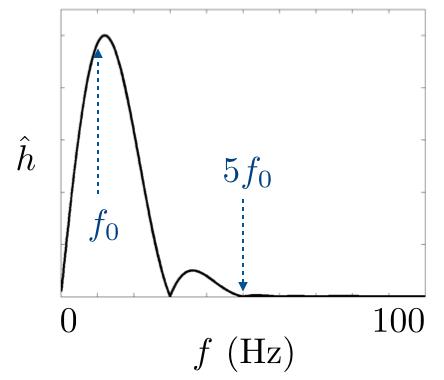}

\small (a) \hspace{5cm} (b)
\caption{\label{FigSource} (a) Temporal signal at a source point (\ref{JKPS}) for $f_0=10$ Hz, and (b) corresponding spectrum.}
\end{figure}
}
\end{itemize}


\subsection{Validation of the immersed interface method for large interface}\label{SecExpPlane}

To begin with, we examine how accurately the immersed interface method discretizes the homogenized jump conditions. To that aim,  we consider  a plane wave at oblique incidence  $\theta_I$ on the interface. This case allows  for an exact, or say reference,  solution of the homogenized problem
in the frequency domain  (see \ref{SecExact})
\beq\label{ty}
\bU(x,y,\omega)=
\bU_I(x,y,\omega)+R(\omega)\bU_R(x,y,\omega)+T(\omega)\bU_T(x,y,\omega),
\eeq
with $\bU_{I,R,T}$ given by (\ref{PlaneWaves}) and $(R,T)$ by (\ref{RT}).
 Afterwards the solution $\bU(x,y,t)$ in the time domain (called {\em reference homogenized solution} in the following) is deduced by  discrete inverse Fourier transform of $\bU(x,y,\omega)$. 
  
Numerically, this solution has to be recovered by imposing, at each time step, the reference  solution $\bU(i\Delta x, j\Delta x,n\Delta t)$ on the 2 lines ($i=0,1$ and $(N_x-1)$, $N_x$) and the 2 columns ($j=0,1$ and $(N_y-1)$, $N_y$) at the edges of the computational domain. Once these boundary conditions have been imposed, the numerical scheme has to be able to produce the solution in the whole domain. This numerical  solution  is referred as the {\em numerical homogenized solution} in the following. 

\vspace{.3cm}

In the simulations, we used a temporal signal with Fourier dependence  given by (\ref{Fourier}) at the central frequency $f_0=10$ Hz, resulting in a wave packet with  the central and smallest wavelengths of 150 m and  30 m respectively. The homogenized interface mimicking the array of Neumann rectangles (with spacing 20 m, filling fraction 0.5 and thickness 20 m) is thus 20 m large, and associated to the interface parameters (\ref{param}). The Fourier synthesis is done using  $N_f=256$ modes with a uniform frequency step $\Delta f=0.4$ Hz around $f_0$ and yields the reference solution $\bU$ imposed at the 2 points boundaries of the computational domain, as previously described. The computational domain is 1200 m$\times$ 1200 m large and it is discretized using $\Delta x=\Delta y=2$ m ($N_x=N_y=600$), thus from the CFL condition $\Delta t=0.0013$ s.


\subsubsection{Plane wave at normal incidence}
We first consider  a  normal incidence, for which $\bU$ is independent of $x$. 
Fig. \ref{FigPlaneTheta0} shows the pressure field of the reference solution which is imposed in the whole domain at the initial time  $t=0$; this initial time has been chosen before the  wave hits the interface (the profile of this one-dimensional field is also reported). 
\begin{figure}[htbp]
\label{FigPlaneTheta0}
\centering
\includegraphics[height=.39\columnwidth]{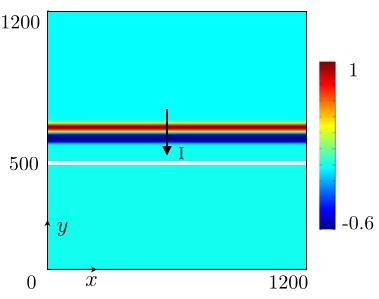}
\includegraphics[height=.39\columnwidth]{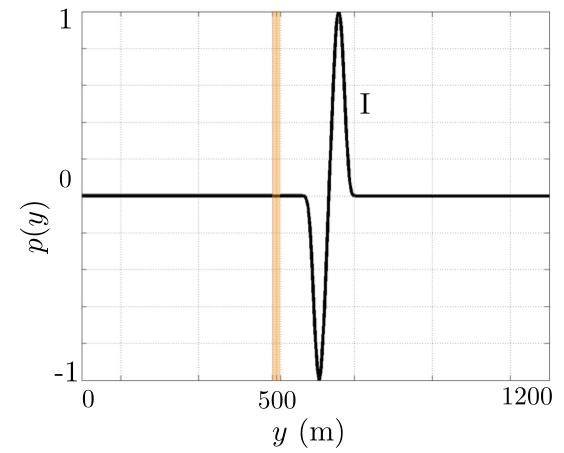}

\small (a) \hspace{6.5cm} (b)

\caption{\label{FigPlaneTheta0} (a) Pressure field of the incident wave (for the reference solution) imposed at the initial time of the simulation and (b) corresponding $y$- profile.  }
\end{figure}

\begin{figure}[htbp]
\centering
\includegraphics[height=.4\columnwidth]{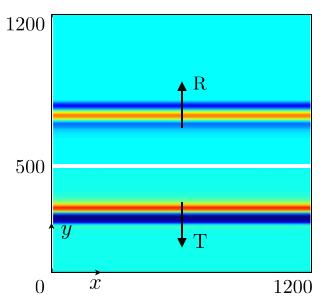}
\includegraphics[height=.4\columnwidth]{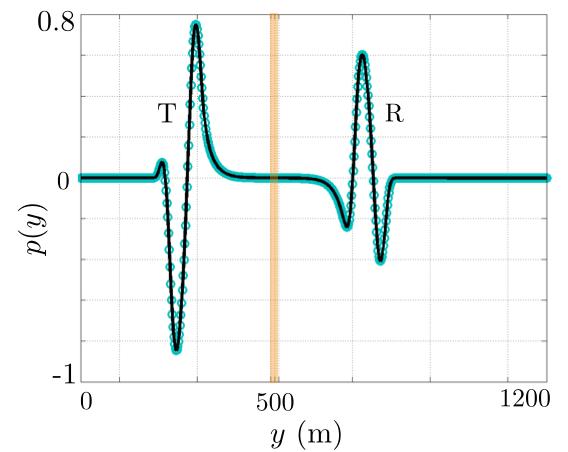}

\small (a) \hspace{6.5cm} (b)

\caption{\label{FigPlaneTheta02}  (a) Pressure field of the numerical homogenized solution at $t=0.2$ s (see text) and (b)  Pressure  profiles of the  reference   (black line) and numerical   (open symbols) homogenized solutions; the relative error between both profiles is of 0.2\%}
\end{figure}

The reference and the numerical homogenized solutions are then computed in time;  the pressure fields of the numerical solution after 158 iterations ($t=0.2$ s) is reported in Fig. \ref{FigPlaneTheta02}(a). 
Fig. \ref{FigPlaneTheta02}(b) reports the profiles along $y$ of the two solutions. The  discrepancy  between both solutions is 0.2\%,  a very low error which can be  attributable to the discretization.  We conclude that the immersed interface method discretized correctly the jump conditions.

\vspace{.3cm}

\begin{figure}[htbp]
\centering
\includegraphics[height=.38\columnwidth]{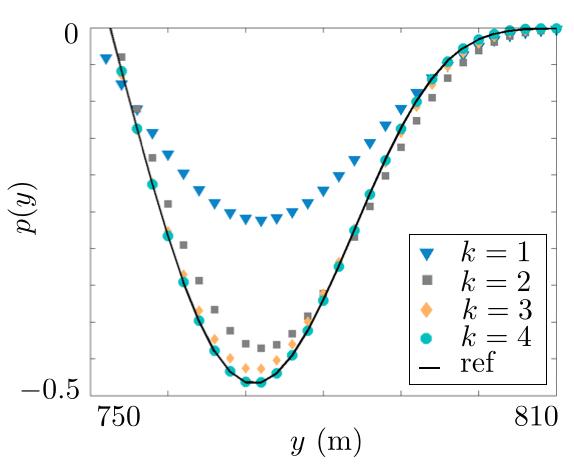}
\includegraphics[height=.38\columnwidth]{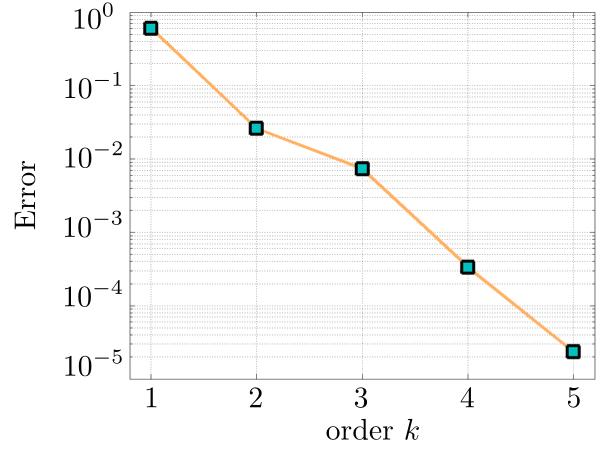}

\small (a) \hspace{6.5cm} (b)

\caption{\label{FigErreurK}  Influence of the order $k$ 
in the computation of the numerical homogenized solution. (a) Zoom on the pressure profile of Fig. \ref{FigPlaneTheta02}(b); the profiles for $k=1,2,3$ and 4 are shown (the profiles for $k=4$ and 5 are indiscernable). (b) Errors  of the numerical solution compared to the reference solution (calculated for $y\in[0, 1200]$) as a function of the order $k$.}
\end{figure}

To further inspect the accuracy of the method, we report results varying the $k$-values and the spatial discretization. With the same $\Delta x=2$ m, we used the orders $k=1$ to 5  to compute the numerical homogenized solution.  Fig.  \ref{FigErreurK}(a) reports the resulting pressure profiles 
along $y$  at $t=0.2$ s (the profile for $k=5$ is indiscernible of the one for $k=4$ and it is not 
reported).  Fig.  \ref{FigErreurK}(b) reports the errors between these profiles and their reference counterpart as a function of $k$. While the solution computed with $k=1$ misses the correct order of magnitude of the reference homogenized solution, good results are obtained for $k>1$, and the error becomes incidental for $k=5$ as used later on.

\begin{figure}[h!]
\centering
\includegraphics[height=.45\columnwidth]{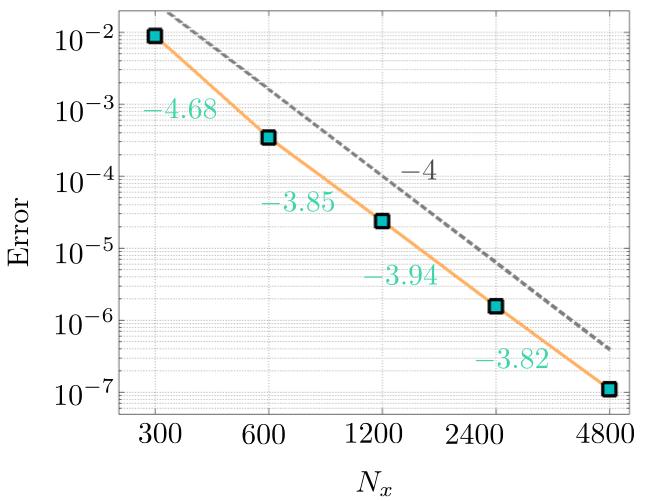}
\caption{\label{FigErreurN} Error as a function of the mesh size $N_x$  (for  $k=5$). The numerical values indicate the slopes between two successive points, the slope -4 is shown in dotted line.}
\end{figure}

Next, we used $k=5 $ and various discretizations for $N_x=300$ up to $N_x=4800$. The  error   is reported in Fig. \ref{FigErreurN} as a function of $N_x$. 

\noindent  The order measured (being the slope of the curve) is close to 4 at this order as soon as  $N_x=300$ ($\Delta x=4$ m) and this corresponds to the  best convergence that we can expect since the order 4 is the  order of ADER 4 in homogeneous medium. It is worth noting that this order 4 is not found for $k<5$, and such high order has a numerical cost. However, 
the difference in accuracy between $k=4$ and  $k=5$ being very small, a good compromise in practice is to choose $k=4$.
 In terms of the spatial resolution, we have said that the smallest wavelength is about 30 m; with an error less than 1\textperthousand   $\;$ for $N_x=600$ ($\Delta x= 2$ m), we can estimate that 15 grid nodes  per wavelength calculated for the smallest wavelength is a good criterion to fix the spatial resolution (and this conclusion holds for $k=4$ as well).
 

\subsubsection{Plane wave at oblique incidence and tilted interface}

The same comparison between the reference and computed homogenized solutions is performed in the case of a wave at oblique incidence and, more importantly, using a tilted interface. As in the previous section, the reference homogenized solution is obtained by discrete inverse Fourier transform of (\ref{ty}) with $\theta_I$ the angle between the incident wave and the tilted interface  (see \ref{SecExact}).  The case of a tilted interface allows us to inspect another aspect of the efficiency of the immersed interface method, namely its capacity to account for the interface shape with a subcell resolution on a Cartesian grid. In the present case, this means that the real slope of the interface is accounted for, instead of a crude  stair-step discretization.

\begin{figure}[htbp]

\centering
\includegraphics[height=.41\columnwidth]{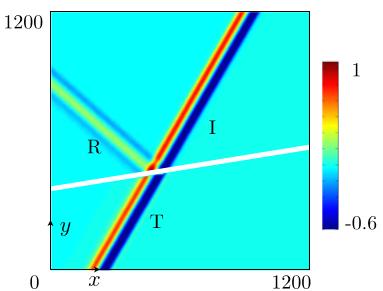}
\includegraphics[height=.41\columnwidth]{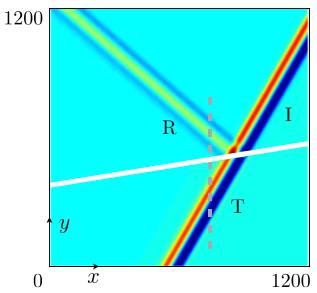}
\\

\small (a) \hspace{6cm} (b) 

\caption{\label{FigTheta10-Carte} (a) Pressure field of the numerical homogenized solution (identical to the reference one) imposed at the initial time of the simulation, (b) Pressure field of the numerical homogenized solution at $t=0.2$ s; the vertical slice (dotted lines) at $x=740$ m, $y\in[100,800]$ m is used in Fig.  \ref{FigTheta10-Coupe}.}
\end{figure}

We considered the interface with a tilt angle of about 10$^\circ$  with  $Ox$  and the incident  wave packet  makes an angle $-30^\circ$  with $Ox$. Figs. \ref{FigTheta10-Carte} show the pressure fields of the numerical homogenized  solution at the initial time (identical to the reference one) and  after the wave packet has propagated  ($t=0.2$ s corresponding to 158 iterations). As expected,   no spurious diffractions have appeared. 


\begin{figure}[htbp]

\centering
\includegraphics[width=.53\columnwidth]{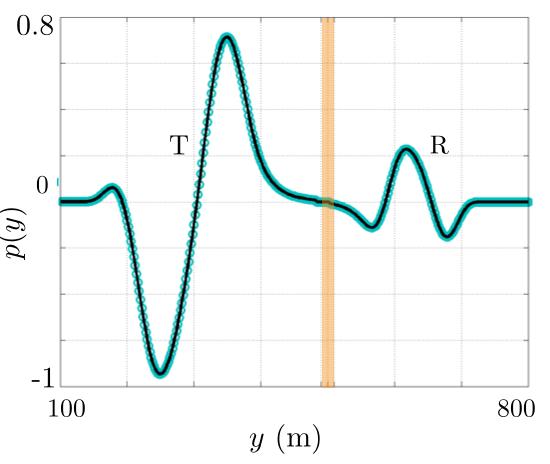}

\caption{\label{FigTheta10-Coupe} Pressure profiles along the slice in dotted line in Fig.  \ref{FigTheta10-Carte}(b), of the reference  (black line) and  numerical  (open symbols) homogenized solutions; the agreement between both is 0.5\textperthousand.}
\end{figure}

More quantitatively, the pressure profile along the vertical slice  ($x=740$ m and $y\in[100, 800]$ m) is reported in  Fig. \ref{FigTheta10-Coupe} together with the corresponding reference solution (the scattered wave packets (R and T) are visible on these profiles). The discrepancy  between the two profiles is of  0.5\textperthousand,  as small as in the case of a non tilted interface. This accuracy could not be obtained with a piecewise constant approximation of the interface shape, and it confirms that the slope of interface is accurately accounted for in the numerical scheme.

\subsection{Validation of the homogenized problem in the time domain}\label{SecExpDirect}

In the previous section, we have inspected the ability of our immersed interface method to properly account for the homogenized jump conditions (\ref{JCa}). Another question is whether or not the homogenized problem (\ref{Acoustic}) is a good approximation of the real   one (\ref{AcousticD}); this question is addressed  now through two examples. First, we consider an  array of scatterers  along a straight line $\Gamma$, which corresponds to the  configuration for which the jump conditions have been derived \cite{Marigo16b}. Next, we consider an array located onto   a curved line $\Gamma$, for which we extended heuristically the  jump conditions. This allows us to inspect the intuitive idea that such extension is possible for small curvature of $\Gamma$. 

\vspace{.4cm}

The real problem is solved numerically  following the method presented in   \cite{Lombard08}.  The numerical method uses  a  scheme ADER 4, and is able to accurately  account for free boundaries at an interface by affecting so-called fictitious values of the solution inside the sound-hard scatterers. In the spirit, these fictitious values are the equivalent of the modified values presented in this paper, and used in the homogenized thick interface. 


\subsubsection{The case of a straight array of sound- hard scatterers}

The array of rectangular scatterers is placed at $y=500$ m along the $x$-axis. Each rectangle is 10 m large along $x$, with spacing $h=$20 m, and $e=$20 m thick along $y$. In the homogenized problem, the interface is $e=$20 m thick (the region $y=[490,510] $ m is not resolved), and associated to the interface parameters (\ref{param}). We considered  a source point at   $(x_s=600,y_s=620)$ m emitting the short pulse $h(t)$ in (\ref{JKPS}), with central frequency $f_0$. 
When not specified, we used a fine grid $\Delta x=0.125$ m for the real problem and a coarse grid $\Delta x=2$ m for the homogenized problem (in both cases, $k=4$ has been considered). These are the mesh sizes required  to get converged solutions (from $\Delta x=2$ m, reducing the mesh size to $\Delta x=$ 0.5 m produces about 20\% variations of the solution of the real problem,  while  the solutions of the homogenized  have already converged, with variations less than 0.1\%).  
\begin{figure}[t!]
\begin{center}
\begin{tabular}{cc}
\includegraphics[width=.9\columnwidth]{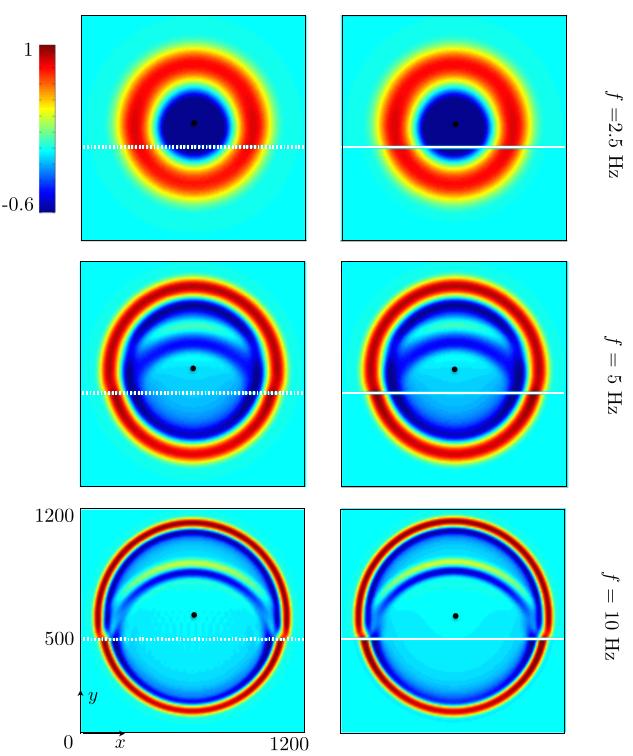} \\

\small (a) $\pr(x,y,t=0.38$ s) \hspace{2cm} (b) $p(x,y,t=0.38$ s) 

\end{tabular}
\caption{\label{FigDirectCartes} Pressure fields  $\pr$ (real problem with the array) and $p$ (homogenized problem) computed with ADER 4 at $t=0.38$ s for  $f_0=2.5$,  5 and 10 Hz. $f_0$ is the central frequency of the signal $h(t)$ in (\ref{JKPS}) imposed by the source point (indicated by the black point). }
\end{center}
\end{figure}

\vspace{.2cm}  
 
\begin{figure}[h!] 
\begin{center}
\begin{tabular}{cc}
\includegraphics[height=.35\columnwidth]{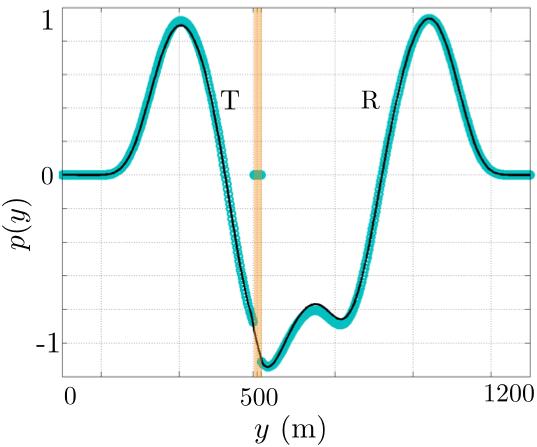} \hspace{.5cm}
\includegraphics[height=.35\columnwidth]{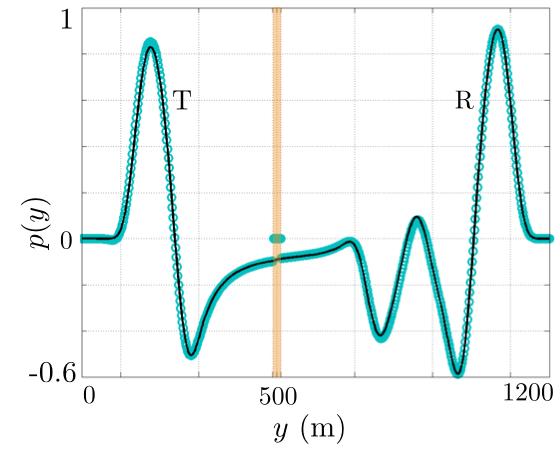} \\

\small (a) $f_0=2.5 $ Hz\hspace{4cm} (b) $f_0=5 $ Hz
\\

\\

\includegraphics[height=.35\columnwidth]{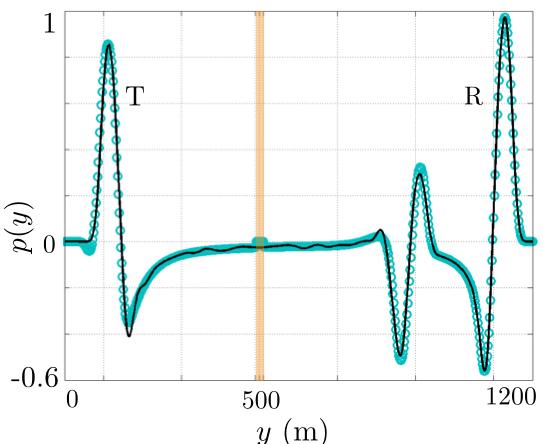}  \hspace{.5cm}
\includegraphics[height=.35\columnwidth]{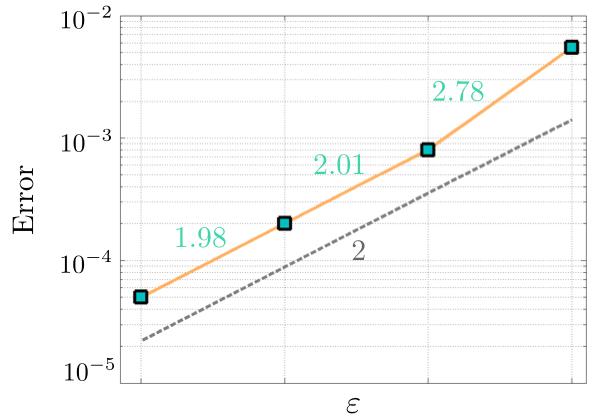} 
\\

(c) $f_0=10 $ Hz\hspace{4cm} (d)

\end{tabular}

\caption{\label{FigDirectZoom1} (a-c) Pressure profiles along the centerline  $x=600$ m from Figs. \ref{FigDirectCartes},  $\pr$  (blue symbols) and $p$ (black lines). (d) Error $|\pr -p|/|p|$ calculated on the transmitted wave $y\in [0 490]$ m as a function of $\varepsilon=2\pi f_0/c$ (an extra point at $f_0=20$ Hz has been considered); dotted grey line shows the $\varepsilon^2$ law. }
\end{center}
\end{figure}

The pressure fields $\pr$ computed in the real problem and $p$ in the homogenized one are reported in Figs. \ref{FigDirectCartes} for central frequencies $f_0=2.5$, 5 and 10 Hz, at $t=0.38$ s (corresponding profiles along the centerline $x=600$ m are reported in Figs. \ref{FigDirectZoom1}). The discrepancy  between the two fields  is of about 5\% for $f_0=$ 2.5 and 5 Hz, and it is of 10\% for $f_0=10$ Hz, and these orders of magnitude are in agreement with those reported in the frequency regime, see \cite{Marigo16a}  (we measured the discrepancy  between both fields outside the thick interface $y\in [490, 510]$ m where $p$ is not defined). Note that, at $f_0=10$ Hz, the central and smallest wavelengths are 150 and 30 m, leading to $\varepsilon \sim 0.8- 4$, so overcoming the intuitive limit  $\varepsilon=1$ for  the validity of the homogenization. We report in Fig. \ref{FigDirectZoom1}(d) the relative errors   for increasing $\varepsilon$-value. Here, the error has been calculated on the part of the profiles corresponding to the transmitted wave packet $y\in [0, 490]$ m (thus avoiding to cross the interface) and with $\varepsilon=2\pi f_0 h/c$ (an extra point at $f_0=20$ Hz has been added). The observed $\varepsilon^2$ scaling is consistant with the expected second-order accuracy of the interface homogenization model. 

\vspace{.4cm}

The computations have been performed using a fine grid $\Delta x=0.125$ m for the real problem and a coarse grid $\Delta x=2$ m for the homogenized problem; as previously said, we checked that these meshes are required to get converged solutions. This is not obvious at $f_0=10$ Hz where  the source generates wavelengths  of the same order of magnitude than the size of the array; thus, we could expect that the two problems, real and homogenized, require the same mesh size, but we observe that it is not the case. Heuristically, this can be explained as follows. The fine grid needed for the real problem  resolves the smallest  scale, and it turns out that this smallest scale is associated to  the evanescent field, excited  in the vicinity of the array (this small scale is visible in Fig. \ref{FigDirectCartes}(a) for $f_0=10$ Hz). This means that the usual rule of say 15 nodes per wavelength has to apply to this near field scale  and not only to the incident wavelength; however, the near field scale is not known {\em a priori} and it depends on how deeply the evanescent field is excited, so it is difficult to anticipate how fine has to be the grid   (a  discussion of this point is presented in \cite{Marigo16d}).

In the homogenized problem, the near field effect is encapsulated in the interface parameters and this is possible since the near field is essentially a static field. Thus,  the mesh size is limited by the usual rule on the incident field only. This is confirmed here; with a minimum incident wavelength of 30 m, a mesh size $\Delta x=2$ m is sufficient to get a converged  homogenized solution; next the validity of the field depends on how close the real near field is close to the static one, and this is lost progressively by increasing the frequency.  

\begin{figure}[h!]
\begin{center}
\begin{tabular}{cc}
\includegraphics[width=.6\columnwidth]{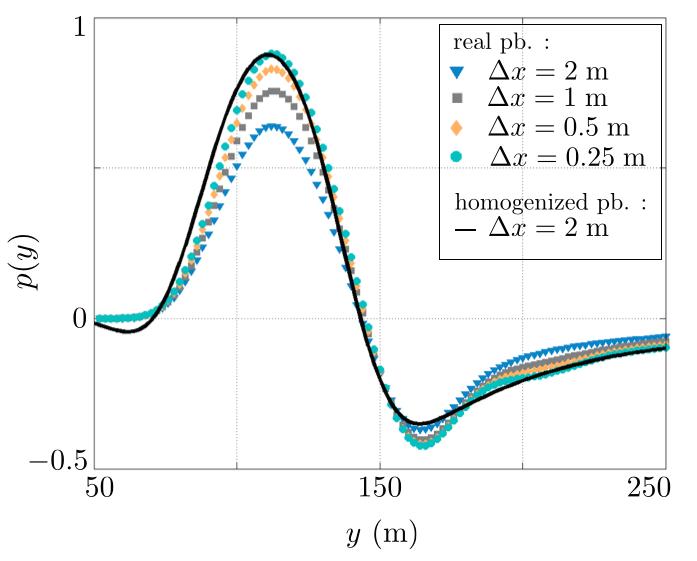} \\

\end{tabular}

\caption{Zoom on the profiles of the solution in the real problem (symbols) compared to the homogenized solution (black plain line) for $f=$ 10 Hz (same profile as in Fig. \ref{FigDirectZoom1}); in the real problem, $\Delta x$ has been reduced from 2 m (as used in the homogenized problem) to 0.25 m.}
\label{FigDirectZoom2} .
\end{center}
\end{figure}

Now, let us inspect how different is the story for the mesh size needed in the real problem.  Fig. \ref{FigDirectZoom2} shows a zoom of the profiles  along the centerline $x=0$ ($y\in [50, 250]$ m) for decreasing mesh size $\Delta x$.  It is visible that the solution of the real problem continues  to converge toward the solution of the homogenized problem (up to the error due to model) for $\Delta x$ well beyond $\Delta x=$ 2 m, and it has not converged before $\Delta x=0.25$ m. This means that the evanescent field experiences rapid variations with a typical exponential decrease over lengths of  few meters ($\Delta x=0.25$ m solves, as a rule of the thumb, variations over 4 meters of the evanescent field, to be compared to the 30 meters wavelength). 

In conclusion, the gain in replacing  the real problem by the homogenized one at low frequency  is simply given by the ratio  between the geometrical size of the array ($h,e$) on the  wavelength. Besides, the smaller is the frequency, the better is the agreement between the homogenized and the real problem. It is obviously for these small $\varepsilon$ values that the homogenized problem can mimic the real problem in the most efficient way.
 
The situation is more involved for intermediate frequencies (say $\varepsilon$ of order unity). In this regime, the  homogenized solution becomes less efficient to describe the real problem. Nevertheless, the gain in the numerical cost remains important; this is because the smallest scale to be resolved in the real problem is associated to the near field variation, and not anymore to the array size. Because higher frequency produces stronger scattering, the evanescent field may contain scales significantly smaller than the array size. For these higher frequencies, a compromise between the accuracy of the homogenized solution and the numerical gain has to be found, and this  depends on the wanted precision.  As an indication in the  numerical gain at $f_0=10$ Hz, a computational time 
of 1 minute for the homogenized problem corresponds to a computational time of 10 hours in the real problem. The extra time needed in the real problem is not only due to the smaller mesh grid, but also to the smaller time step imposed by the CFL condition. 
  

\subsection{Variable homogenized interface}\label{SeExpVar}

\begin{figure}[htbp]
\begin{center}
\begin{tabular}{cc}
\includegraphics[scale=0.35]{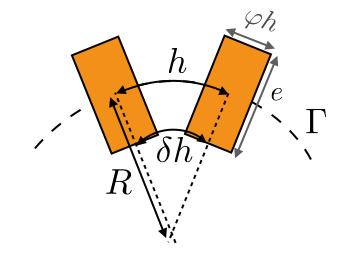} 
\end{tabular}
\vspace{-0.4cm}
\caption{\label{FigCourbure} Effect of the local curvature on the relative positions of two scatterers. $R$ is the local radius of curvature, producing a smallest distance $\delta h\leq h$ between the two scatterers (and $\delta h=h$ for $1/R=0$).}
\end{center}
\end{figure}

As previously said, the numerical implementation of the jump conditions (\ref{JCa}) have been extended  to the case of a curved line $\Gamma$. Below,  we  report results varying the curvatures. For small curvatures, this allows us to validate the numerical implementation of the jump conditions along a curved line; also, we inspect   the error due to the model when  increasing the curvature. For high curvatures, we expect the jump conditions to be modified. It is outside the scope of the present paper to derive such conditions, but let us estimate the maximum curvature below which  we expect the jump conditions (\ref{JCa}) to be unaffected. The parameters $(B,C_1,C_2)$ have been calculated in  static problems to account for the boundary layers effects near the scatterers; if these boundary layers are significantly modified because of  the local curvature, the parameters will be affected as well; it is worth noting that, if the case, the homogenized problem becomes more tricky since the parameters will vary along $\Gamma$ (if the curvature varies, that is for any curve $\Gamma$ except a circle). Fig. \ref{FigCourbure} shows the relative position of two -rectangular- scatterers for a local curvature $1/R$, producing a minimum distance of $\delta h$ between them (and $\delta h=h$ for infinite $R$, or zero curvature). 

\noindent The modification of the boundary layers in the static problems involved to calculate $(B,C_1,C_2)$ are expected to depend of how  $\delta h/h$ is far from 1. Note also the condition $\delta h>\varphi h$ for no overlapping of the scatterers. It is easy to see that this leads to
\beq\label{critere}
1\geq \frac{\delta h}{h}=\frac{R-e/2}{R}> \varphi.
\eeq
Now, in addition to the condition of  no-overlapping which  imposes $e/R<2 (1-\varphi)$, small values of $e/R$-values (or small $1-\delta h/h$) are expected to ensure the validity of (\ref{JCa}), and this is what we shall inspect further.

\vspace{.3cm}

To easily increase the curvature, we consider a curved line $\Gamma$ in the form of a sinusoid with mean value at $y_m=$ 500 m and $y_\Gamma= y_m+A\cos 2\pi (x-x_s)/D$ ($x_s=600$ m). We kept a fixed $A=$ 10 m value and varied  $D$ from $D=$ 1000 m and 125 m. The corresponding maximum local curvatures are given by $\kappa= A(2\pi/D)^2$, leading to a minimum local radius of curvature $R$ from $R=$ 2500 m to 40 m. This smallest value of $R$ corresponds to twice the minimum value imposed by non overlapping (\ref{critere}).   

\begin{figure}[h!]
\begin{center}
\begin{tabular}{cc}
\includegraphics[scale=0.42]{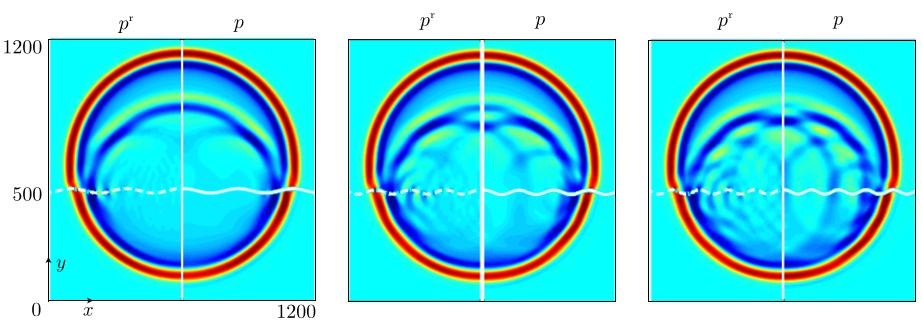} 
\end{tabular}
\vspace{-0.2cm}
\end{center}
\small \hspace{1cm} (a) \quad $1-\delta h/h=0.06$ \hspace{1.7cm} (b)\quad $1-\delta h/h=0.15$ 
\hspace{1.7cm} (c)\quad  $1-\delta h/h=0.25$ 
\caption{\label{FigCarpette2} Wavefields for a curved mean line $\Gamma$. Expect the shape of $\Gamma$, the calculations are identical to those of Fig. \ref{FigDirectCartes} for $f_0=$ 10 Hz. On (a-c), the left panels show the wavefields $p^r$ in the real problem (with dotted white lines indicating $\Gamma$); the right panels the wavefields $p$ in the homogenized problem (with the white region indicating the homogenized interface. }
\end{figure}

\begin{figure}[h!]
\centering
\includegraphics[width=.45\columnwidth]{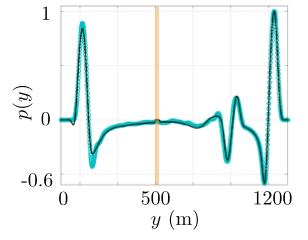} 
\includegraphics[width=.45\columnwidth]{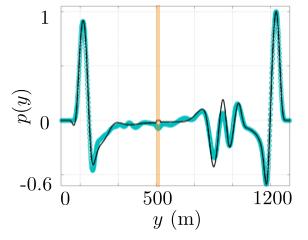} 
\includegraphics[width=.45\columnwidth]{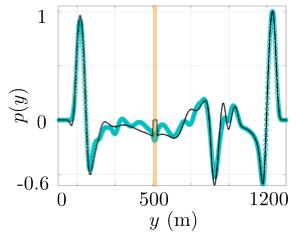} 
\includegraphics[width=.45\columnwidth]{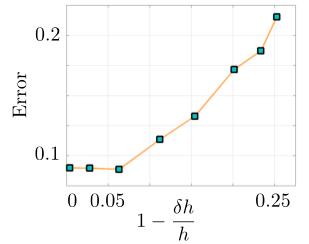} 
\caption{\label{FigZoomCarpette2}  (a-c) Pressure profiles along the centerline $x=600$ m from Figs. \ref{FigCarpette2},  $\pr$  (blue symbols) and $p$ (black lines). (d) Error $|\pr -p|/|p|$ as a function of $(1-\delta h/h)$, Eq. (\ref{critere}). }
\end{figure}

To begin with, we report in Figs. \ref{FigCarpette2} and \ref{FigZoomCarpette2}(a-c) the wavefields and the  profiles on the centerline $x=600$ m in the real problem and in the homogenized problem for $f_0=$ 10 Hz. 
The reported  time is $t=0.38$ s, and the calculations have been performed in the same conditions as  in Figs. \ref{FigDirectCartes}-\ref{FigDirectZoom1}). Here, we have considered $D=$ 250 m ($R=$ 160 m), $D=$ 160 m ($R=$ 65 m) and $D=$ 125 m ($R=$ 40 m). Corresponding  values of $(1-\delta h/h)$ are 0.06, 0.15 and 0.25 respectively, for a maximum allowed value of 0.5. By comparison of the results in Figs. \ref{FigCarpette2}-\ref{FigZoomCarpette2}  with those obtained with a straight line $\Gamma$ in Figs. \ref{FigDirectCartes}-\ref{FigDirectZoom1}, it is visible that even a small curvature produces significant modification in the field pattern. This is particularly noticeable regarding the signal often referred as the  "coda", which corresponds to the signal between the two main wavefronts directly transmitted and reflected by the interface. In the real problem, increasing the curvature of $\Gamma$ enhances multiple scattering effect in the region of the scatterers, which feeds the coda region. Also noticeable is the fact that the homogenized solutions reasonably reproduce the main features of this coda region, although they are unable to reproduce its finest scales.

Finally,  Fig. \ref{FigZoomCarpette2}(d) shows the error as a function of $(1-\delta h/h)$ where additional $D$ values have been considered. The result is as expected: for $\delta h\simeq h$, the effect of the curvature  is incidental and the error remains the same as for a straight line $\Gamma$; increasing further the curvature produces an increase in the error. In the presented case, this happens for $(1-\delta h/h)$ above  10\%, 
thus for a relative minimum distance between the scatterers 5 times smaller than the minimum distance of overlapping.


\section{Conclusion}\label{SecCon}

We have proposed a numerical method to implement   jump conditions established within a homogenization approach. These jump conditions  allow us to replace the real problem of the wave propagation though  an array of sound hard scatterers by an equivalent homogenized problem, much simpler to handle numerically. The numerical method has been validated considering a  scattering problem for which an explicit solution is available, and comparisons of the solutions in the homogenized problem and in the real problem have confirmed the ability of the homogenized problem to advantageously replace the real one. Advantageously is meant with a computational time highly reduced; in the cases reported here, the computational time is typically reduced by a factor 500.  In fact, and as previously said, accounting for the jump conditions requires  additional calculations being time independent. Thus, they  are done once and for all independently of the wave source and before the time resolution is considered. Then, the time resolution is not more demanding than for a problem of wave propagation in a homogeneous medium.

In terms of the possible extensions of the numerical method, several directions seems  of interest to us. The first is rather incremental. We have considered thicknesses of the scatterers being small but sufficiently large with respect to the expected mesh size; this is because we assumed that points of certain stencils may fall within the homogenized interface (which is not resolved). If the scatterer thickness passes below  the  mesh size, the region of the interface will always been contained between two nodes; this case has  been considered already, notably for vanishing scatterer thickness \cite{AnneSo} (the harmonic regime is considered in this reference) and it is strictly more simple to handle numerically. Nevertheless, because of the practical interest in such thin arrays (the gain in the numerical implementation is even more significant), this has to be done. The second extension is numerically not so demanding but it requires to adapt the homogenization approach. The jump conditions have been established for the scatterers being located along a straight line and we have inspected their possible extension to the case of a curved line. This has confirmed the intuitive idea that large local curvatures require a modified version of the jump conditions. It is worth noting that this would lead to a more involved homogenized problem since the curvature being defined locally along the mean line, the parameters entering in the jump conditions would be local as well. Nevertheless, and again in regard with the practical interest of such configurations, a generalization to scatterers located onto curved lines deserves interest. 
 
Finally, several extensions concern the nature of the scatterers, and let us mention two situations which are not trivial extensions of the present work. The first concerns  scatterers associated to Dirichlet 
boundary conditions: it is typically  metallic arrays illuminated by a polarized electromagnetic  wave in the far infrared regime. In this case, effective boundary conditions at each side of the interface have to be considered, rather than  jump conditions  \cite{Marigo16a}. The second is a bolder extension of the present work. It concerns scatterers with material properties having  high  contrasts with respect  to the surrounding medium, such that resonances inside the scatterers are possible. In this case, the jump conditions cannot be derived in the time domain. The calculations are done in the harmonic regime, revealing  interface parameters being frequency dependent. Thus, the numerical implementation of  these jump conditions in the time domain requires to handle these frequency dependent parameters. 
 

\appendix
\section{Interface parameters $(B,C_1,C_2)$ for rectangular sound-hard scatterers}
\label{SecIntPar}
The interface parameters involved in the jump conditions (\ref{JCa}) have been calculated for 
rectangular  sound-hard scatterers in \cite{Marigo16b,Marigo16d}. As they are written in (\ref{JCa}), they have the dimension of lengths and read
\beq
\begin{array}{ll}
 \ds B=
\frac{e}{1-\varphi}+\frac{2}{\pi}\log\left(\sin \frac{\pi(1-\varphi)}{2}\right)^{-1},\\ 
 \ds C_1=e(1-\varphi),\\ 
\end{array}\eeq
and, as a rule of the thumb for the last parameter
\beq\left\{
\begin{array}{ll}
 \ds C_2\simeq e(1-\varphi)-\frac{\pi}{8}(1-\varphi)^2, & \textup{if this leads to a positive value},\\
\ds C_2\simeq 0, & \textup{otherwise},\\
\end{array}
\right.
\eeq
for rectangular scatterers being of length $\varphi h$ with spacing $h$ and of thickness $e$.
A more accurate of $C_2$ can be obtained by solving a so-called elementary problem and a simple script to do so is provided in \cite{Marigo16b}.


\section{Scattering of a plane wave at oblique incidence on a plane homogenized interface}

\label{SecExact}

\begin{figure}[htbp]
\centering
\includegraphics[width=.495\columnwidth]{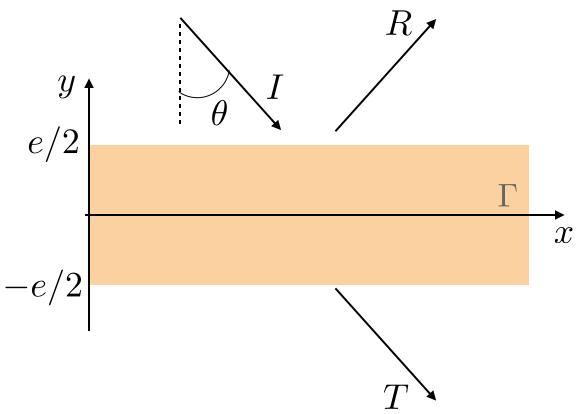} 
\caption{\label{FigOndePlane} Incident plane wave (I) impacting a plane homogenized interface, leading to a reflected (R) and a transmitted (T) plane waves.}
\end{figure}
  
We consider a plane wave at oblique incidence $\theta$ on the thick interface and the problem to solve is (\ref{Acoustic}). We want to determine $\bU(x,y,\omega)$ in (\ref{ty}), and $\bU$ is defined in  (\ref{UAB}). Below, we shall calculate the pressure $p$ afterwards $\bU=(v_x,v_y,p)^T$ is deduced using  (\ref{Acoustic}) written in the harmonic regime, with time dependence $e^{-i\omega t}$, whence
\beq\left\{\begin{array}{l}
\ds v_x(x,y,\omega)=-\frac{i}{\rho \omega}\; \frac{\partial p}{\partial x}(x,y,\omega),\\[6pt]
\ds v_y(x,y,\omega)=-\frac{i}{\rho \omega}\; \frac{\partial p}{\partial y}(x,y,\omega).\\
\end{array}\right.\label{rrr}\eeq

For this one dimensional problem, the pressure field (\ref{UAB}) reads
\begin{equation}
p(x,y,\omega)= e^{ i\omega x\sin \theta/c} \times 
\left\{
\begin{array}{ll}
e^{-i \omega (y-e/2)\cos \theta/c} + R
e^{i \omega (y-e/2)\cos \theta/c}, & x >e/2,\\
T e^{-i \omega (y+e/2)\cos \theta/c}, & x <e/2,\\
\end{array}\right.\label{PlaneWaves}
\end{equation}
and we want to determine the scattering  coefficients $(R,T)$.
It is sufficient to  inject (\ref{PlaneWaves}) in the jump conditions (\ref{JCa}) using (\ref{Acoustic}), and setting
the impedance $Z=\rho\,c\,\cos \theta$ and the parameters $(\alpha,\beta)$ 
\begin{equation}
\begin{array}{l}
\ds \alpha=\frac{\rho}{2}\left(C_{1}\,\cos^2 \theta+C_{2}\,\sin^2 \theta \right),\quad
\ds \beta=\frac{ B}{2\,c}\; \cos \theta,
\end{array}
\end{equation}
to get the system satisfied by $R$ and $T$ 
\begin{equation}
\left\{
\begin{array}{l}
\ds\ds
(T-R)\left(1+i\omega\,\beta\right)=\left(1-i\omega\,\beta\right),\\[12pt]
(T+R)\left(Z-i\omega\,\alpha\right)= \left(Z+i\omega\,\alpha\right).
\end{array}
\right.
\label{SystRT}
\end{equation}
We get
\begin{equation}
R=\frac{i\omega\,(\alpha+Z\beta)}{(Z+i\omega \alpha)(1+i\omega \beta)},\hspace{0.5cm}
T=\frac{Z-\omega^2\,\alpha\beta}{(Z+i\omega \alpha)(1+i\omega \beta)},
\label{RT}
\end{equation}
from which $|R|^2+|T|^2=1$. These expressions of $(R,T)$, together with (\ref{PlaneWaves}) and (\ref{rrr}) give the reference solution $\bU(x,y,\omega)$ used in (\ref{ty}). 

\vspace{3cm} 



\begin{thebibliography}{99}

\bibitem{AnneSo}
{\sc A.S. Bonnet-Bendhia, D. Drissi, N. Gmati}, 
{\em Simulation of muffler's transmission losses by a homogenized finite element 
method}, J. Comp. Acoust., 12(3) (2004) 1-28.

\bibitem{Capdeville1}
{\sc Y. Capdeville  and J.-J. Marigo},
	 {\em Second-order homogenization of the elastic wave equation for non-periodic layered media},
	 {Geophys. J. Int.},
{170} (2007) {823--838}.

\bibitem{Capdeville2}
{\sc Y. Capdeville, L. Guillot, J.-J. Marigo},
	 {\em 2-D non-periodic homogenization to upscale elastic media for p-sv waves},
{Geophys. J. Int.},
	 {182} (2010) {903-922}.

\bibitem{Cioranescu} {\sc D. Cioranescu, P. Donato}, {\em  An introduction to homogenization}, volume 17 of Oxford Lecture Series in Mathematics and its Applications. The Clarendon Press Oxford University Press, New York, 4, 118.
ISO 690	 (1999).

\bibitem{Delourme1}
{\sc B. Delourme, H. Haddar, P. Joly}, {\em Approximate models for wave propagation across thin  periodic interfaces}, J. Math. Pures Appl., 98 (2012) 28-71. 

\bibitem{Delourme2}
{\sc B. Delourme}, {\em High-order asymptotics for the electromagnetic scattering by thin periodic layers},  Math. Meth. Appl. Sciences, 38(5)  (2015) 811-833.

\bibitem{Chiavassa11}
{\sc G. Chiavassa, B. Lombard}, {\em Time domain numerical modeling of wave propagation in 2D heterogeneous porous media}, J. Comput. Phys., 230 (2011), 5288-5309. 

\bibitem{Gustafsson75}
{\sc B. Gustafsson}, {\em The convergence rate for difference approximations to mixed initial boundary value problems}, Math. Comput., 29-130 (1975), 396-406.

\bibitem{Li94}
{\sc Z. Li and R.~J. LeVeque}, {\em The Immersed Interface Method for elliptic equations with discontinuous coefficients and singular sources}, SIAM J. Num. Anal., 31 (1994), 1019-1044.

\bibitem{Lombard04}
{\sc B. Lombard, J. Piraux}, {\em Numerical treatment of two-dimensional interfaces for acoustic and elastic waves}, J. Comput. Phys., 195-1 (2004), 90-116.

\bibitem{Lombard06}
{\sc B. Lombard, J. Piraux}, {\em Numerical modeling of elastic waves across imperfect contacts}, SIAM J. Scient. Comput., 28-1 (2006), 172-205. 

\bibitem{Lombard08}
{\sc B. Lombard, J. Piraux, C. G\'elis, J. Virieux}, {\em Free and smooth boundaries in 2-D finite-difference schemes for transient elastic waves}, Geophys. J. Int., 172 (2008), 252-261. 

\bibitem{Lorcher05}
{\sc F. L\"orcher, C. Munz}, {\em Lax-Wendroff-type schemes of arbitrary order in several space dimensions}, IMA J. Numer. Anal., (2005), 1-28.

\bibitem{Marigo16a}
{\sc J.~J. Marigo, A. Maurel}, {\em Two scale homogenization to determine effective parameters of thin metallic structured films}, to appear in Proc. R. Soc. A (2016). 

\bibitem{Marigo16c}
{\sc A. Maurel, J.~J. Marigo, A. Ourir},  {\em Homogenization of ultrathin metallo-dielectric structures leading to transmission conditions at an equivalent interface}, J. Opt. Soc. Am. B, 33(5), 947-956 (2016). 

\bibitem{Marigo16b}
{\sc J.~J. Marigo, A. Maurel}, {\em Homogenization models for thin rigid structured surfaces and films}, J. Acoust. Soc. Am. 140(1)  (2016), 260-273.

\bibitem{Marigo16d}
{\sc J.~J. Marigo, A. Maurel}, {\em An interface model for homogenization of acoustic metafilms}, submitted  (2016).
Available at
{\tt https://www.researchgate.net/profile/Agnes$\_$Maurel2}.

\bibitem{Marigo11}
{\sc J.~J. Marigo, C. Pideri}, {\em The effective behavior of elastic bodies containing microcracks or microholes localized on a surface}, Int. J. Damage. Mech., 20 (2011), 1151-1177.

\bibitem{Pendry04}
{\sc J.~B. Pendry, L. Martin-Moreno, F.~J. Garcia-Vidal}, {\em Mimicking surface plasmons with structured surfaces}, Science, 5685 (2004), 847-848.

\bibitem{Press92}
{\sc W.~H. Press, S.~A. Teukolskyn, W.~T. Vetterling, B.~P. Flannery}, {\it Numerical Recipes in C: The Art of Scientific Computing}, Cambridge University Press (1992). 

\bibitem{Schwartzkopff04}
{\sc T. Schwartzkopff, M. Dumbser, C. Munz}, {\em Fast high-order ADER schemes for linear hyperbolic equations}, J. Comput. Phys., 197-2 (2004), 532-539.
%
\bibitem{Strickwerda99}
{\sc J.~C. Strikwerda}, {\it Finite Difference Schemes and Partial Differential Equations}, Chapman \& Hall (1999).  

\bibitem{Zhang97}
{\sc C. Zhang and R.~J. LeVeque}, {\em The Immersed Interface Method for acoustic wave equations with discontinuous coefficients}, Wave Motion, 25 (1997), 237--263.  

\end{thebibliography}
\end{document}